\documentclass[applsci,article,submit,moreauthors,pdftex,10pt,a4paper]{mdpi} 
\pdfoutput=1
\usepackage[T1]{fontenc}
\usepackage[utf8]{inputenc}
\usepackage{bm}
\usepackage{amsmath}
\usepackage{amssymb}
\usepackage{esint}
\usepackage{graphicx}
\preto{\abstractkeywords}{\nolinenumbers}

\firstpage{1} 
\articlenumber{x}
\doinum{10.3390/------}
\pubvolume{xx}
\pubyear{2017}
\copyrightyear{2017}
\externaleditor{Academic Editor: name}
\history{Received: date; Accepted: date; Published: date}

\Title{Impact of graphene on the polarizability of a neighbour nanoparticle: a dyadic Green's function study}

\Author{Bruno Amorim $^{1,\ddagger}$\orcidA{}, P. A. D. Gon\c{c}alves $^{2,3}$\orcidB{},  Mikhail I. 
Vasilevskiy  $^{4,}$\orcidC{}, and N. M. R. Peres$^{4,\ddagger}$*\orcidD{}}

\AuthorNames{Bruno Amorim, P.A.D. Gonçalves, Mikhail I. Vasilevskiy,  and N. M. R. Peres}

\address{%
$^{1}$ \quad Department of Physics and CeFEMA, Instituto Superior T\'{e}cnico,
University of Lisbon, Av. Rovisco Pais, PT-1049-001 Lisboa, Portugal; amorim.bac@gmail.com\\
$^{2}$ \quad Department of Photonics Engineering and Center for Nanostructured Graphene, Technical University
of Denmark, DK-2800 Kgs. Lyngby, Denmark; padgo@fotonik.dtu.dk\\
$^{3}$ \quad Center for Nano Optics, University of Southern Denmark, DK-5230 Odense M, Denmark\\
$^{4}$ \quad Department and Centre of Physics, and QuantaLab, University
of Minho, Campus of Gualtar, PT-4710-374, Braga, Portugal; mikhail@fisica.uminho.pt, peres@fisica.uminho.pt}

\corres{Correspondence: peres@fisica.uminho.pt; Tel.: +351 253604334}

\secondnote{These authors contributed equally to this work.}

\abstract{
	We discuss the renormalization of the polarizability of a nanoparticle in the
    presence of either (i) a continuous graphene sheet or (ii) a plasmonic graphene grating, taking into account
    retardation effects.
	Our analysis demonstrates that the excitation of surface plasmon-polaritons in graphene
	produces a large enhancement of the real and imaginary parts of the renormalized
	polarizability. We show that the imaginary part can be changed by a factor of up to 100 relatively to its value
	in the absence of graphene.
	We also show that the resonance in the  case of the grating is narrower
	than in the continuous sheet. In the case of the grating it is shown
	that the resonance can be tuned by changing the grating geometric parameters.}


\begin{document}

\setcounter{section}{0} 

\section{Introduction}

The polarizability of a nanoparticle is a response function which relates the electric dipole 
moment produced in it to an externally applied eletric field.  The polarizability is not an intrinsic property of the nanoparticle, but it actually 
depends on the environment  which it is embedded in \cite{Novotny,Pelton_book, Amendola_2017_review}. As such, a 
nanoparticle's polarizability will be modified by the presence of an underlying substrate. The study of this 
problem is of significant interest, since in most experimental setups the nanoparticle (NP) is 
placed directly onto a dielectric substrate or at a given distance from it. In previous 
studies in which the radiation scattered by a dielectric NP has been measured using dark-field microscopy, it 
has been shown that the presence of the substrate leads to a redshift of the NP's resonance with 
respect to the situation where the NP is in vacuum~\cite{nl2012_Bozh,scirep12,renormalized-polarizability-2}.

The polarizability of a nanoparticle at a given frequency is a complex quantity, with its real and imaginary 
parts describing, respectively, the reactive and dissipative responses of a nanoparticle subjected to
an electromagnetic field. Therefore, the imaginary part of the polarizability  controls the
extinction and absorption cross-sections of a nanoparticle subjected to an impinging electromagnetic field 
\cite{William_Barnes,Review-on-nanoparticles} (see also Appendix \ref{sec:Derivation-rate}). These quantities 
are essential for the understanding of scattering experiments of electromagnetic radiation involving 
nanoparticles, either isolated or forming clusters. In particular, the former case 
has been a topic of much interest in the context of single-molecule or single-particle spectroscopies~\cite{Link:2010,Link:2011}. 
The knowledge of the imaginary part of the polarizability 
is also essential in order to understand the phenomena of blackbody and thermal friction experienced by a 
neutral 
nanoparticle in close proximity to an interface between two media \cite{imaginary-part}. It is therefore of major importance to 
understand how the imaginary part of the polarizability is renormalized relatively to its value in vacuum 
when it is near an interface, the most common setup in experiments. 

It would be of particular relevance, from the device engineering viewpoint, if the dielectric properties
of the interface, near which the nanoparticle is located, could be tuned. This would provide a route for 
controlling the value of the  nanoparticle polarizability in real time. Such approach
is not viable when we consider the interface between two conventional dielectrics or between a metal and a dielectric, 
since the interface has fixed properties by construction. Fortunately, there is a possible and technologically feasible route 
to overcome this limitation. Adding a graphene sheet between an interface involving two different dielectrics 
provides an additional degree of freedom to the problem. Indeed the Fermi energy of a graphene sheet can be
controlled in real time using a gate. Tuning the Fermi energy of graphene changes the local dielectric 
environment around the nanoparticle and therefore the value of the imaginary part of the nanoparticle 
polarizability. This is the opportunity we will explore in this paper.

Incidentally, the problem of  nanoparticle's polarizability renormalization in the presence of a substrate 
is also relevant for the characterization the dielectric properties of a scanning near-field
optical microscope (SNOM). SNOM is a technique frequently used to image and characterize surface polaritons in 
graphene \cite{imaging_plasmons_2012} and other two-dimensional materials,
such as boron nitride \cite{Dai_hBNpolariton}. More recently, exciton-polaritons have also been studied 
in layered transition metal dichalcogenides using the same method~\cite{hu2017imaging}. Indeed the SNOM tip
is frequently modeled as a dipole, as is the nanoparticle \cite{SNOM-GREEN}. Therefore, understanding how a 
nanoparticle changes its dielectric properties under illumination allows us to also understand the problem
of SNOM tip illuminated with THz radiation during the excitation of surface polaritons in graphene and other 
two-dimensional materials.

\begin{figure}
\centering
\includegraphics[width=9cm]{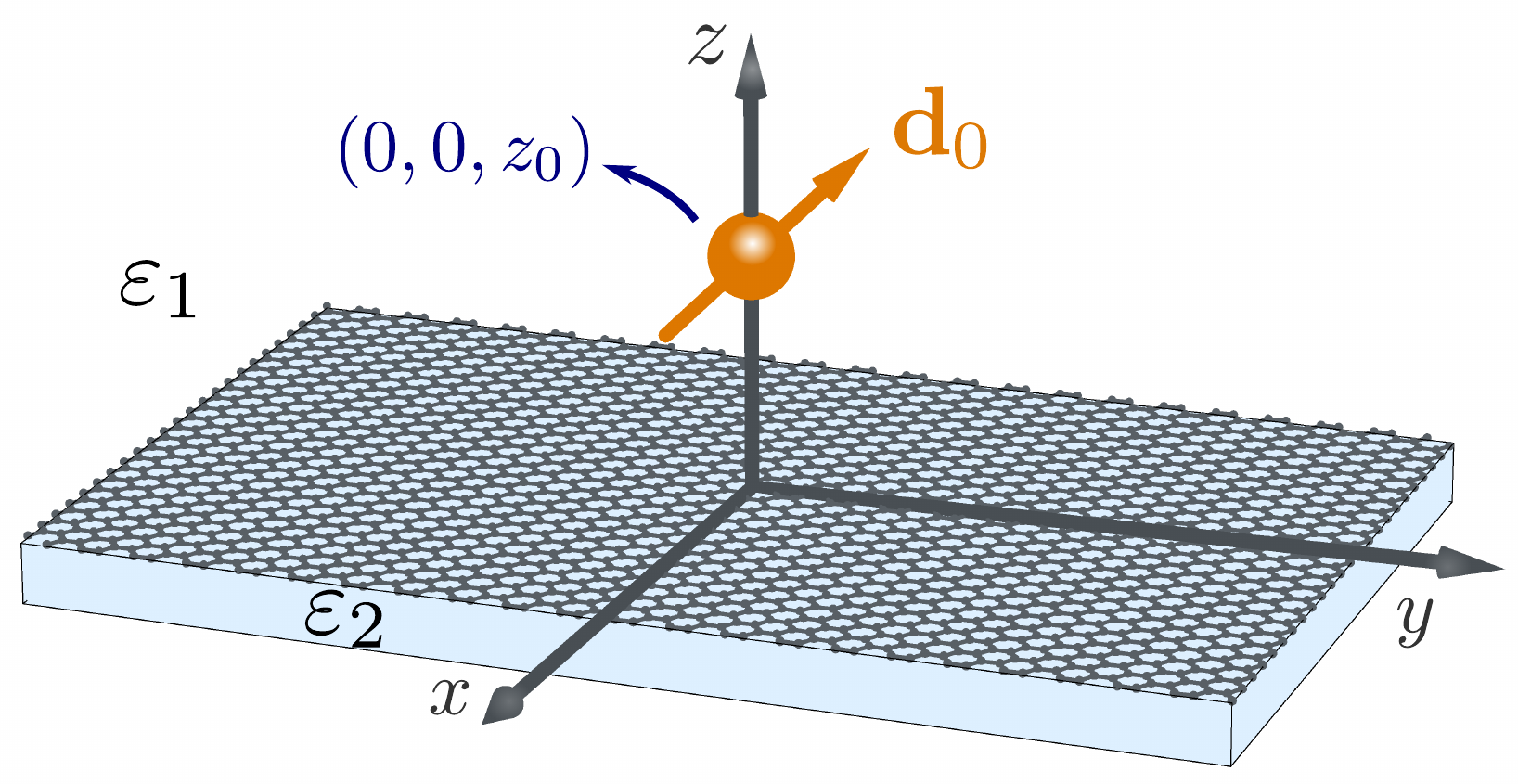}
\includegraphics[width=9cm]{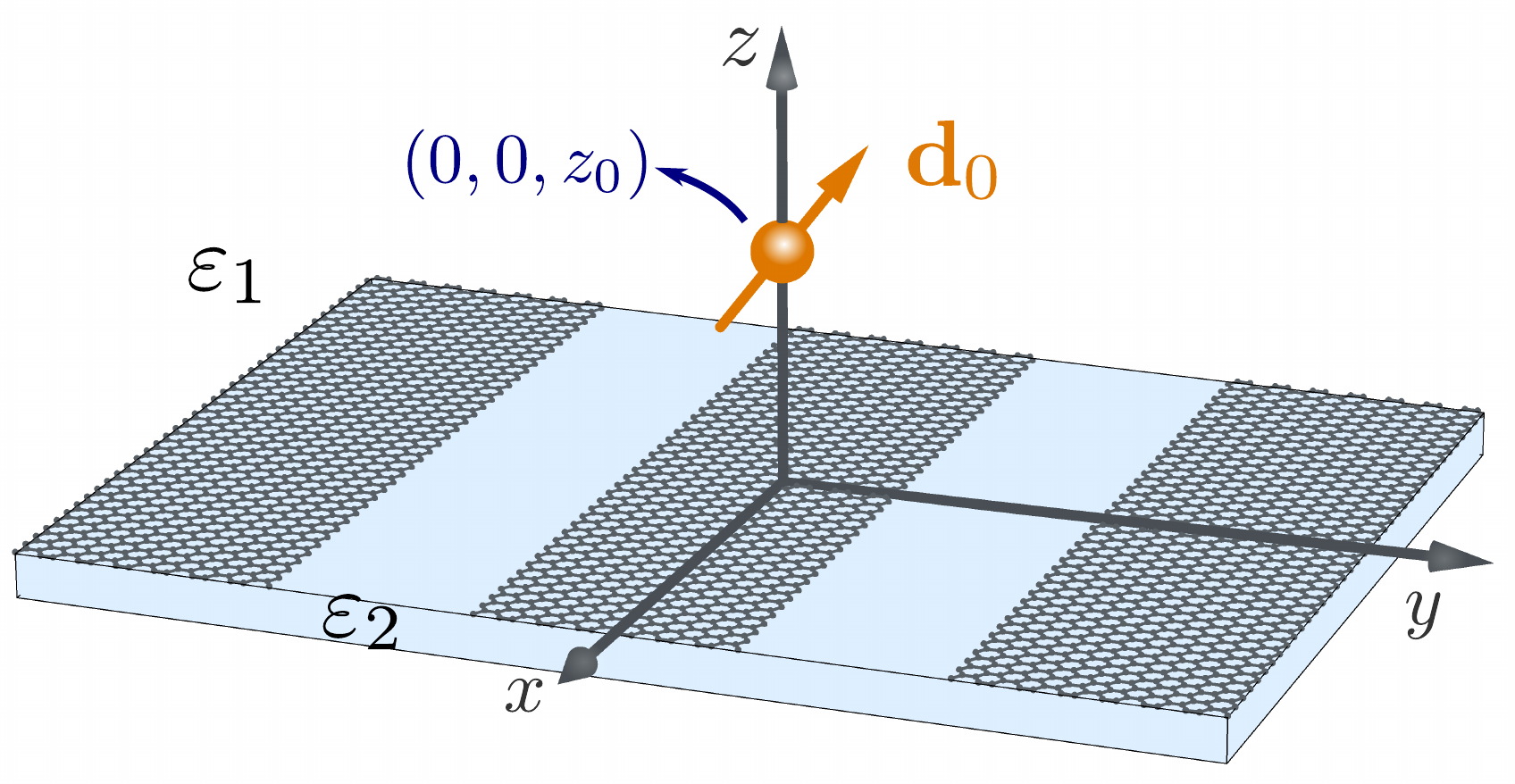}
\caption{\label{fig:System}The two systems considered in this paper: a graphene sheet (top) and a 
graphene-grid of ribbons (bottom) located in between two dielectrics. A nanoparticle is located at position 
$\mathbf{r}_0=(0,0,z_0)$ and is characterized by a polarizability tensor $\alpha_0$  in 
vacuum. In addition, a plane wave impinges on the nanoparticle and on graphene coming from $z=+\infty$.}
\end{figure}

In this work, we study either the renormalization of a nanoparticle polarizability located near the interface 
between two dieletrics interspaced with a doped graphene sheet, or with an array of graphene ribbons 
(see Figure~\ref{fig:System}). One of the dielectrics is the vacuum and the other acts as substrate
for the support of the graphene sheet. In order to keep the analysis simple we shall restrict ourselves to the case of 
a non-dispersive and non-dissipative substrate, characterized by a frequency independent and real dielectric
constant. We explore the  imaginary part of the polarizability in the THz range of the 
electromagnetic spectrum, a spectral region where graphene supports surface plasmon-polaritons \cite{BLUDOV:2013aa,Sorger-THZ,book}. As
we will see, the excitation of these polaritons leads to a significant change of the polarizability of both a 
metallic and a semiconductor nanoparticles. Indeed, the bare polarizability  of a metallic nanoparticle in 
vacuum is essentially constant in the THz with a very small imaginary part of the polarizability.
However when located near a graphene sheet the polarizability undergoes a strong renormalization, specially 
in what concerns its imaginary part. 

Although the problem of modeling the polarizability of a nanoparticle close to a graphene sheet has been 
considered before by some of the authors of the present paper \cite{Jaime}, that work relied on a 
electrostatic approximation. The present work goes beyond that, taking into account retardation 
effects, allowing us to correctly describe the imaginary part of the polarizability. It should be noted that 
the problem of determining the nanoparticle's polarizability in the presence of a homogeneous flat dielectric 
substrate has also been considered previously both in the electrostatic approximation 
\cite{polarizability_no_graphene} and in with the full electrodynamic approach 
\cite{renormalized_polarizability,renormalized-polarizability-2,Dahan-2012}. 

The goal of this work is fourfold: (i) to extend the 
study of Ref. \cite{Jaime}  including retardation effects, thus using a more general formalism; (ii) to 
bring together in a single paper a formalism that is scattered in the literacture using many different 
notations; (iii) to introduce a rigorous formulation of the dyadic Green's function formalism that is absent 
in many papers; and (iv) to extend this approach to the case where a nanoparticle has both  dipolar electric and 
dipolar magnetic moments. 

This paper is organized as follows: in Section \ref{sec:dyadic-GF}
we introduce the concept of dyadic Green's function for the electric field as a
tool to obtain the electric field in the presence of source currents. In Section
\ref{subsec:Free-space} we study in detail the electric field dyadic Green's function in free-space (or in a 
homogeneous medium). The Weyl's, or angular spectrum, representation 
of the dyadic Green's function is introduced in Section \ref{subsec:Weyl}. This representation is well 
suited to deal with the problem of radiation scattering at planar interfaces. It is
also shown that the dyadic Green's functions can be expressed in terms of the tensor product of the electric 
field $s$-- and $p$--polarization vectors. In Section \ref{subsec:Source-and-scattered}, we focus on the 
problem of scattering at a planar interface and define the reflected and transmitted Green's functions. In 
Section \ref{sec:Renormalization-polarizability} we deduce the polarizability of a nanoparticle close to an 
interface covered by graphene. We start  defining 
and studying the polarizability  of a nanoparticle embedded in vacuum, in Section 
\ref{subsec:def-polarizability}. The approach is generalized in Section 
\ref{subsec:renormalization-interface} to the case of a nanoparticle close to a planar interface. This 
general description is then used to describe the renormalization of a nanoparticle's polarizability close to a 
continuous graphene sheet and to a graphene grating in Sections \ref{subsec:Renormalized-graphene-sheet} and 
\ref{subsec:Renormalized-graphene-grating}.   
In  Section \ref{sec:Inclusion-of-magnetic} we present a generalization of the formalism to the 
case where the nanoparticle has both electric and magnetic dipole moments. Such a magnetic moment can be 
generated, even for nanoparticles formed by a non-magnetic material, due to induced currents inside the 
nanoparticle \cite{magnetic-dipole}, and can actually be the main contribution for the polarizability in 
the case of dielectric NPs~\cite{nl2012_Bozh,scirep12,renormalized-polarizability-2}.
Finally, a set of Appendices provides some auxiliary results.

\section{Dyadic Green's function for the electric field}\label{sec:dyadic-GF}

\subsection{Free-space dyadic Green's function}\label{subsec:Free-space}

The goal of this section is to introduce the dyadic Green's function
that allows us to solve the wave equation for the electric field in the presence of source currents. Although the
material in this section is relatively well known, we present it here
in some detail both for the sake of completeness and to fix notation used throughout the paper. The 
inhomogeneous wave equation for the electric field reads (see Appendix \ref{sec:Derivation-of-wave_equation})
\begin{equation}
\nabla\times\nabla\times\mathbf{E}(\mathbf{r},\omega)-\frac{\omega^{2}}{v_{n}^{2}}\mathbf{E}(\mathbf{r},
\omega)=i\omega\mu_{n}\mu_{0}\mathbf{j}_{f}(\mathbf{r},\omega),\label{eq:wave_eq_Electric}
\end{equation}
where $v_{n}=1/\sqrt{\epsilon_{0}\epsilon_{n}\mu_{0}\mu_{n}}$ is the
speed of light in a medium with the relative permittivity and permeability
given, respectively, by $\epsilon_{n}$ and $\mu_{n}$, and $\mathbf{j}_{f}(\mathbf{r},\omega)$
is the free current not taken into account by $\epsilon_{n}$ and
$\mu_{n}$. For future use we also define $k_{n}=\omega/v_{n}$. The
electric field free-space dyadic Green's function, 
$\overleftrightarrow{G_0}(\mathbf{r},\mathbf{r}^{\prime},\omega)$,
is introduced in order to solve the inhomogeneous Eq.~(\ref{eq:wave_eq_Electric})
in integral form as
\begin{equation}
\mathbf{E}(\mathbf{r},\omega)=\mathbf{E}_{0}(\mathbf{r},\omega)+i\omega\mu_{n}\mu_{0}\int 
d\mathbf{r}^{\prime}\overleftrightarrow{G_0}(\mathbf{r},\mathbf{r}^{\prime},\omega)\mathbf{\cdot 
j}_{f}(\mathbf{r^{\prime}},\omega),\label{eq:electric_field}
\end{equation}
where $\mathbf{E}_0(\mathbf{r},\omega)$ is a solution of the homogeneous wave equation (that is, in the 
absence of free currents, $\mathbf{j}_{f}(\mathbf{r},\omega)$).
For a current due to a point dipole located at $\mathbf{r}=\mathbf{r}_{0}$,
we have 
$\mathbf{j}_{f}(\mathbf{r},\omega)=-i\omega\mathbf{d}_{0}\delta\left(\mathbf{r}-\mathbf{r}_{0}\right)$, where 
$\mathbf{d}_0$ is the electric dipole moment. In this case, Eq.~(\ref{eq:electric_field}) reduces to (for 
$\mathbf{r}\neq\mathbf{r}_0$)
\begin{equation}
\mathbf{E}(\mathbf{r},\omega)=\mathbf{E}_{0}(\mathbf{r},\omega)+\omega^{2}\mu_{n}\mu_{0}\overleftrightarrow{G}
_{0}(\mathbf{r},\mathbf{r}_{0},\omega)\cdot\mathbf{d}_{0}.
\end{equation}
We must now determine $\overleftrightarrow{G_0}(\mathbf{r},\mathbf{r}^{\prime},\omega)$.
In the standard Green's function approach, one would attempt
to compute $\overleftrightarrow{G_0}(\mathbf{r},\mathbf{r}^{\prime},\omega)$
by solving \cite{Novotny}

\begin{equation}
\nabla\times\nabla\times\overleftrightarrow{G_0}(\mathbf{r},\mathbf{r}^{\prime},\omega)-k_{n}^{2}
\overleftrightarrow{G_0}(\mathbf{r},\mathbf{r}^{\prime},\omega)=\overleftrightarrow{I}\delta(\mathbf{r}
-\mathbf{r}^{\prime}),\label{eq:GF_eq}
\end{equation}
where $\overleftrightarrow{I}$ is the $3\times3$ identity matrix.
Such equation is, apparently, easily solved writing the 
Green's function in Fourier components, reducing the above equation to an algebraic
equation, whose solution is obtained by inverting a $3\times3$ matrix .
However, difficulties arise when one tries to invert the Fourier transform back to real space, particularly 
in obtaining the correct behaviour of the Green's function for $\mathbf{r}=\mathbf{r}^{\prime}$,
which will be essential in the next sections. 

We will, therefore, pursue an alternative approach in order to determine $\overleftrightarrow{G}_0$, which 
follows the method originally described in Ref.~\cite{Yaghjian_1980}. The first step to 
determine $\overleftrightarrow{G}_0$ in this approach is noticing that the inhomogeneous wave equation for the 
electric field can be written as an inhomogeneous Helmholtz equation (see 
Appendix~\ref{sec:Derivation-of-wave_equation} for a derivation)
\begin{equation}
-\nabla^{2}\mathbf{E}(\mathbf{r},\omega)-k_{n}^{2}\mathbf{E}(\mathbf{r},\omega)=i\omega\mu_{n}\mu_{0}\left[
\mathbf{j}_{f}(\mathbf{r},\omega)+\frac{v_{n}^{2}}{\omega^{2}}\nabla\left(\nabla\cdot\mathbf{j}_{f}(\mathbf{r}
,\omega)\right)\right].
\end{equation}
The general solution of the Helmholtz equation can be written as (see Appendix~\ref{sec:GF_Helmholtz})
\begin{equation}
\mathbf{E}(\mathbf{r},\omega)=\mathbf{E}_{0}(\mathbf{r},\omega)+i\omega\mu_{n}\mu_{0}\int_{\backslash 
V_{\delta}(\mathbf{r})}d^{3}\mathbf{r}^{\prime}g_{0}\left(\mathbf{r},\mathbf{r}^{\prime},
\omega\right)\left [ 
\overleftrightarrow{I}+\frac{1}{k_{n}^{2}}\nabla^{\prime}\nabla^{\prime}\right]\mathbf{j}_{f}(\mathbf{r}^{ 
\prime},\omega),\label{eq:integral_Helmholtz_Efield}
\end{equation}
where 
\begin{equation}
g_{0}\left(\mathbf{r},\mathbf{r}^{\prime},\omega\right)=\frac{e^{ik_{n}\left|\mathbf{r}-\mathbf{r}^{\prime}
\right|}}{4\pi\left|\mathbf{r}-\mathbf{r}^{\prime}\right|},\label{eq:Helmholtz_greenfunction}
\end{equation}
is the Green's function for the scalar Helmholtz equation \cite{Yaghjian_1980,Duffy,Novotny}, and 
$\int_{\backslash 
V_{\delta}(\mathbf{r})}$
represents integration in the principal value sense, where an infinitesimal
volume, $V_{\delta}(\mathbf{r})$, enclosing the point $\mathbf{r}^{\prime}=\mathbf{r}$ is excluded.
We have written $\nabla^{\prime}\nabla^{\prime}\equiv\nabla^{\prime}\otimes\nabla^{\prime}$
with $\otimes$ denoting the tensor product and the prime indicates that the derivative is over the 
$\mathbf{r}^{\prime}$ variables. The Helmholtz Green's function
is the solution of
\begin{equation}
\left[-\nabla^{2}-k_{n}^{2}\right]g_{0}\left(\mathbf{r},\mathbf{r}^{\prime},\omega\right)=\delta(\mathbf{r}
-\mathbf{r}^{\prime})
\end{equation}
in a way that is clarified in Appendix~\ref{sec:GF_Helmholtz}.
Notice that $g_{0}\left(\mathbf{r},\mathbf{r}^{\prime},\omega\right)$
is integrable, and therefore, the exclusion of the volume $V_{\delta}(\mathbf{r})$
is not usually emphasized. However, it will be important when 
obtaining $\overleftrightarrow{G_0}(\mathbf{r},\mathbf{r}^{\prime},\omega)$. 
Although Eq.~(\ref{eq:integral_Helmholtz_Efield}) already allows to compute the electric field as a 
function of the current, it is useful to obtain an alternative expression which does not involve derivatives 
of the current. Such expression can be obtained by carefully performing integration by parts. It must the 
noticed, that due to the excluded volume surrounding $\mathbf{r}^{\prime}=\mathbf{r}$, boundary
terms are generated during the integration procedure. We obtain
\begin{multline}
\int_{\backslash 
V_{\delta}(\mathbf{r})}d^{3}\mathbf{r}^{\prime}g_{0}\left(\mathbf{r},\mathbf{r}^{\prime},
\omega\right)\nabla^{ \prime}\left(\nabla^{\prime}\cdot\mathbf{j}_{f}(\mathbf{r}^{\prime},\omega)\right)=\\
=-\int_{\partial 
V_{\delta}(\mathbf{r})}d^{2}\mathbf{r}^{\prime}\mathbf{n}^{\prime}\left[g_{0}\left(\mathbf{r},
\mathbf{r}^{ 
\prime},\omega\right)\left(\nabla^{\prime}\cdot\mathbf{j}_{f}(\mathbf{r}^{\prime},\omega)\right)\right]-\int_{ 
\backslash 
V_{\delta}(\mathbf{r})}d^{3}\mathbf{r}^{\prime}\nabla^{\prime}g_{0}\left(\mathbf{r},\mathbf{r}^{
\prime}, \omega\right)\left(\nabla^{\prime}\cdot\mathbf{j}_{f}(\mathbf{r}^{\prime},\omega)\right),
\end{multline}
where $\mathbf{n}^{\prime}$ is a outward pointing unit vector, normal
to the surface $\partial V_{\delta}(\mathbf{r})$ of the enclosing
volume $V_{\delta}(\mathbf{r})$. In the limit of infinitesimal excluded
volume, the first term of the above equation vanishes, since 
the element of area scales as $d^{2}\mathbf{r}^{\prime}\sim\left|\mathbf{r}-\mathbf{r}^{\prime}\right|^{2}$,
while 
$g_{0}\left(\mathbf{r},\mathbf{r}^{\prime},\omega\right)\sim1/\left|\mathbf{r}-\mathbf{r}^{\prime}\right|$.
For the second term, we perform integration by parts once again, obtaining
(for clarity we explicitly write the tensorial components in a Cartesian
basis, with repeated indices being summed over)
\begin{multline}
-\int_{\backslash 
V_{\delta}(\mathbf{r})}d^{3}\mathbf{r}^{\prime}\partial_{i}^{\prime}g_{0}\left(\mathbf{r},\mathbf{r}^{\prime},
\omega\right)\partial_{k}^{\prime}j_{f}^{k}(\mathbf{r}^{\prime},\omega)=\\
=\int_{\partial 
V_{\delta}(\mathbf{r})}d^{2}\mathbf{r}^{\prime}n_{k}^{\prime}\left[\partial_{i}^{\prime}g_{0}\left(\mathbf{r},
\mathbf{r}^{\prime},\omega\right)j_{f}^{k}(\mathbf{r}^{\prime},\omega)\right]+\int_{\backslash 
V_{\delta}(\mathbf{r})}d^{3}\mathbf{r}^{\prime}\left[\partial_{i}^{\prime}\partial_{k}^{\prime}g_{0}
\left(\mathbf{r},\mathbf{r}^{\prime},\omega\right)j_{f}^{k}(\mathbf{r}^{\prime},\omega)\right].
\end{multline}
Now the boundary term is finite. In the limit of an infinitesimal
volume, we take $\mathbf{r}^{\prime}\rightarrow\mathbf{r}$, such
that $j_{f}^{k}(\mathbf{r}^{\prime},\omega)\rightarrow j_{f}^{k}(\mathbf{r},\omega)$
and use the small $\left|\mathbf{r}-\mathbf{r}^{\prime}\right|\rightarrow0$
limit of $\partial_{i}^{\prime}g_{0}\left(\mathbf{r},\mathbf{r}^{\prime},\omega\right)$
Eq.~(\ref{eq:del_g0_limit}). This allows us to write
\begin{multline*}
\int_{\backslash 
V_{\delta}(\mathbf{r})}d^{3}\mathbf{r}^{\prime}g_{0}\left(\mathbf{r},\mathbf{r}^{\prime},\omega\right)\nabla^{
\prime}\left(\nabla^{\prime}\cdot\mathbf{j}_{f}(\mathbf{r}^{\prime},\omega)\right)=\\
=\int_{\backslash 
V_{\delta}(\mathbf{r})}d^{3}\mathbf{r}^{\prime}\nabla^{\prime}\nabla^{\prime}g_{0}\left(\mathbf{r},\mathbf{r}^
{\prime},\omega\right)j_{f}^{k}(\mathbf{r}^{\prime},\omega)-\frac{1}{k_{n}^{2}}\overleftrightarrow{L}_{V_{
\delta}}\cdot\mathbf{j}_{f}(\mathbf{r},\omega)
\end{multline*}
where the dyadic $\overleftrightarrow{L}_{V_{\delta}}$ is defined as \cite{Yaghjian_1980}
\begin{equation}
\overleftrightarrow{L}_{V_{\delta}}=\int_{\partial 
V_{\delta}(\mathbf{r})}\frac{d^{2}\mathbf{r}^{\prime}}{4\pi}\frac{\left(\mathbf{r}^{\prime}-\mathbf{r}
\right)\otimes\mathbf{n}^{\prime}}{\left|\mathbf{r}^{\prime}-\mathbf{r}\right|^{3}},
\label{eq:depolarization_dyadic}
\end{equation}
which can be interpreted as a depolarization term.
Therefore, we can write Eq.~(\ref{eq:integral_Helmholtz_Efield})
as
\begin{align}
\mathbf{E}(\mathbf{r},\omega) & 
=\mathbf{E}_{0}(\mathbf{r},\omega)-i\omega\mu_{n}\mu_{0}\overleftrightarrow{L}_{V_{\delta}}\cdot\mathbf{j}_{f}
(\mathbf{r},\omega)\nonumber \\
 & +i\omega\mu_{n}\mu_{0}\int_{\backslash 
V_{\delta}(\mathbf{r})}d^{3}\mathbf{r}^{\prime}\left[\overleftrightarrow{I}+\frac{1}{k_{n}^{2}}
\nabla\nabla\right]g_{0}\left(\mathbf{r},\mathbf{r}^{\prime},\omega\right)\mathbf{j}_{f}(\mathbf{r}^{\prime},
\omega),
\end{align}
from which we can write $\overleftrightarrow{G_0}(\mathbf{r},\mathbf{r}^{\prime},\omega)$
as
\begin{equation}
\overleftrightarrow{G_0}(\mathbf{r},\mathbf{r}^{\prime},\omega)=\text{P}.\text{V}._{V_{\delta}}\left[
\overleftrightarrow{I}+\frac{1}{k_{n}^{2}}\nabla\nabla\right]g_{0}\left(\mathbf{r},\mathbf{r}^{\prime},
\omega\right)-\frac{1}{k_{n}^{2}}\overleftrightarrow{L}_{V_{\delta}}\delta\left(\mathbf{r}-\mathbf{r}^{\prime}
\right),\label{eq:GO_Dyadic}
\end{equation}
where $\text{P}.\text{V}._{V_{\delta}}$ indicates that the small volume
$V_{\delta}$ centered at $\mathbf{r}^{\prime}=\mathbf{r}$ is to
be excluded. In the standard derivation of $\overleftrightarrow{G_0}(\mathbf{r},\mathbf{r}^{\prime},\omega)$
based on the direct solution of Eq.~(\ref{eq:GF_eq}) it is very easy to miss the 
$\delta\left(\mathbf{r}-\mathbf{r}^{\prime}\right)$ contribution,
which is essential to describe depolarization effects. Notice that
$\overleftrightarrow{L}_{V_{\delta}}$ depends on the shape of the
chosen excluded volume \cite{Yaghjian_1980}. For a sphere it
is straightforward to show that $\overleftrightarrow{L}_{\text{Sphere}_{\delta}}=\overleftrightarrow{I}/3$.
In this case, the free-space dyadic Green's function in real space
can be written as the sum of four terms \cite{Green_in_real_space,delta-identities}
\begin{equation}
\overleftrightarrow{G_0}(\mathbf{r},\mathbf{r}^{\prime},\omega)=\overleftrightarrow{G_0}^{{\rm 
FF}}(\mathbf{r},\mathbf{r}^{\prime},\omega)+\overleftrightarrow{G_0}^{{\rm 
IF}}(\mathbf{r},\mathbf{r}^{\prime},\omega)+\overleftrightarrow{G_0}^{{\rm 
NF}}(\mathbf{r},\mathbf{r}^{\prime},\omega)+\overleftrightarrow{G_0}^{{\rm 
SF}}(\mathbf{r},\mathbf{r}^{\prime},\omega),
\end{equation}
respectively, the far-, intermediate-, near- and self-field terms, which are written as
\begin{align}
\overleftrightarrow{G_0}^{{\rm FF}}(\mathbf{r},\mathbf{r}^{\prime},\omega) & 
=\left(\overleftrightarrow{I}-\hat{\mathbf{R}}\mathbf{\hat{R}}\right)\frac{e^{ik_{n}\vert\mathbf{r}-\mathbf{r}
^{\prime}\vert}}{4\pi\vert\mathbf{r}-\mathbf{r}^{\prime}\vert},\label{eq:FF_GF}\\
\overleftrightarrow{G_0}^{{\rm IF}}(\mathbf{r},\mathbf{r}^{\prime},\omega) & 
=i\left(\overleftrightarrow{I}-3\hat{\mathbf{R}}\mathbf{\hat{R}}\right)\frac{e^{ik_{n}\vert\mathbf{r}-\mathbf{
r}^{\prime}\vert}}{4\pi k_{n}\vert\mathbf{r}-\mathbf{r}^{\prime}\vert^{2}},\label{eq:eq:MF_GF}\\
\overleftrightarrow{G_0}^{{\rm NF}}(\mathbf{r},\mathbf{r}^{\prime},\omega) & 
=-\left(\overleftrightarrow{I}-3\hat{\mathbf{R}}\mathbf{\hat{R}}\right)\frac{e^{ik_{n}\vert\mathbf{r}-\mathbf{
r}^{\prime}\vert}}{4\pi k_{n}^{2}\vert\mathbf{r}-\mathbf{r}^{\prime}\vert^{3}}\label{eq:NF_GF}\\
\overleftrightarrow{G_0}^{{\rm SF}}(\mathbf{r},\mathbf{r}^{\prime},\omega) & 
=-\overleftrightarrow{I}\frac{1}{3k_{n}^{2}}\delta(\mathbf{r}-\mathbf{r}^{\prime}),\label{eq:SF_GF}
\end{align}
where the terms $\overleftrightarrow{G_0}^{{\rm FF}}(\mathbf{r},\mathbf{r}^{\prime},\omega)$,
$\overleftrightarrow{G_0}^{{\rm IF}}(\mathbf{r},\mathbf{r}^{\prime},\omega)$
and $\overleftrightarrow{G_0}^{{\rm NF}}(\mathbf{r},\mathbf{r}^{\prime},\omega)$
are to be understood in the principal value 
sense, and we have introduced the definitions $\hat{\mathbf{R}}=\left(\mathbf{r}-\mathbf{r}^{\prime}\right)/\left|\mathbf{r}-\mathbf{r}^{\prime}
\right|$ and $\hat{\mathbf{R}}\hat{\mathbf{R}}=\hat{\mathbf{R}}\otimes\hat{\mathbf{R}}$.

\subsection{Weyl's or angular spectrum representation of the dyadic Green's function:
an useful formulation for interfaces}\label{subsec:Weyl}

Although Eq.~(\ref{eq:GO_Dyadic}) can be used directly to 
evaluate $\overleftrightarrow{G_0}(\mathbf{r},\mathbf{r}^{\prime},\omega)$,
for many applications such formulation might not be the most useful.
In the the case of scattering by planar interfaces it
is useful to make a (two-dimensioanl) Fourier transform of the fields in the coordinates parallel to
the interface. This representation of the fields and of the Green's 
function is generally referred to as Weyl's
or angular spectrum representation. In this representation, the electric field is written as
\begin{equation}
\mathbf{E}(\mathbf{r},\omega)=\int\frac{d^{2}\mathbf{p}_{\parallel}}{\left(2\pi\right)^{2}}e^{i\mathbf{p}_{
\parallel}\cdot\bm{\rho}}\mathbf{E}(\mathbf{p}_{\parallel},z,\omega),
\label{eq:E_field_Weyl}
\end{equation}
where $\mathbf{p}_{\parallel}$ is the in-plane wave-vector and $\bm{\rho}=\left(x,y\right)$ are in-plane 
coordinates.
In this representation Eq.~(\ref{eq:integral_Helmholtz_Efield})
becomes
\begin{equation}
\mathbf{E}(\mathbf{p}_{\parallel},z,\omega)=\mathbf{E}_{0}(\mathbf{p}_{\parallel},z,\omega)+i\omega\mu_{n}\mu_
{0}\fint 
dz^{\prime}g_{0}\left(\mathbf{p}_{\parallel},z,z^{\prime},\omega\right)\left[\overleftrightarrow{I}-\frac{1}{
k_{n}^{2}}\mathcal{\bm{D}^{\prime}}\mathcal{\bm{D}^{\prime}}\right]\mathbf{j}_{f}(\mathbf{p}_{\parallel},z^{
\prime},\omega),\label{eq:Helmholtz_integral_Weyl}
\end{equation}
where $\mathbf{j}_f(\mathbf{p}_\parallel,z,\omega)$ is the Weyl representation of the current density, 
defined in analogous way to Eq.~(\ref{eq:E_field_Weyl}), 
$\mathcal{\bm{D}^{\prime}}=\mathbf{p}_{\parallel}-i\hat{e}_{z}\partial_{z}^{\prime} $,
$\mathcal{\bm{D}^{\prime}}\mathcal{\bm{D}^{\prime}}\equiv\mathcal{\bm{D}^{\prime}}\otimes\mathcal{\bm{D}^{
\prime}}$, 
$\fint$ represents the principal value integral in one dimension,
excluding the point $z^{\prime}=z$, and $g_{0}\left(\mathbf{p}_{\parallel},z,z^{\prime},\omega\right)$
is the Helmholtz Green's function in the Weyl representation, defined
such that
\begin{equation}
g_{0}\left(\mathbf{r},\mathbf{r}^{\prime},\omega\right)=\int\frac{d^{2}\mathbf{p}_{\parallel}}{
\left(2\pi\right)^{2}}e^{i\mathbf{\mathbf{p}_{\parallel}}\cdot(\bm{\bm{\rho}-\bm{\rho}^{\prime}})}g_{0}
\left(\mathbf{p}_{\parallel},z,z^{\prime},\omega\right).
\end{equation}
The function $g_{0}(\mathbf{p}_{\parallel},z,z^{\prime},\omega)$ can be easily
obtained from the components of the three dimensional Fourier transform of the Helmholtz Green's function,
$g_{0}\left(\mathbf{p},\omega\right)=\left(p_{\parallel}^{2}+p_{z}-k_{n}^{2}\right)^{-1}$, as
\begin{equation}
g_{0}\left(\mathbf{p}_{\parallel},z,z^{\prime},\omega\right)=\int\frac{dp_{z}}{2\pi}e^{ip_{z}\left(z-z^{\prime
}\right)}g_{0}\left(\mathbf{p},\omega\right).
\end{equation}
This integral can be easily performed by contour integration yielding
\begin{equation}
g_{0}\left(\mathbf{p}_{\parallel},z,z^{\prime},\omega\right)=\frac{i}{2\beta_{n}}e^{i\beta_{n}\left|z-z^{
\prime}\right|},\label{eq:Weyl_integral}
\end{equation}
where $\beta_{n}$ is defined as 
\begin{equation}
\beta_{n}=\begin{cases}
\sqrt{k_{n}^{2}-p_{\parallel}^{2}}, & k_{n}^{2}>p_{\parallel}^{2}\\
i\sqrt{p_{\parallel}^{2}-k_{n}^{2}}, & k_{n}^{2}<p_{\parallel}^{2}
\end{cases}.
\end{equation}
Clearly equation (\ref{eq:Weyl_integral}) is written is terms of
both propagating and evanescent waves \cite{Green_decomposition}. Similarly to what we have done
in the previous section, we can rewrite Eq.~(\ref{eq:Helmholtz_integral_Weyl})
by moving the derivatives $\partial_{z}^{\prime}$ from 
$\mathbf{j}_{f}(\mathbf{p}_{\parallel},z^{\prime},\omega)$
to $g_{0}\left(\mathbf{p}_{\parallel},z,z^{\prime},\omega\right)$.
Doing this yields
\begin{equation}
\mathbf{E}(\mathbf{p}_{\parallel},z,\omega)=\mathbf{E}_{0}(\mathbf{p}_{\parallel},z,\omega)+i\omega\mu_{n}\mu_
{0}\int 
dz^{\prime}\overleftrightarrow{G_0}\left(\mathbf{p}_{\parallel},z,z^{\prime},\omega\right)\mathbf{j}_{f}
(\mathbf{p}_{\parallel},z^{\prime},\omega),
\end{equation}
with $\overleftrightarrow{G_0}\left(\mathbf{p}_{\parallel},z,z^{\prime},\omega\right)$ being
the dyadic Green's function in Weyl's representation
\begin{equation}
\overleftrightarrow{G_0}\left(\mathbf{p}_{\parallel},z,z^{\prime},\omega\right)=\text{P.V.}\left[
\overleftrightarrow{I}-\frac{1}{k_{n}^{2}}\mathbf{p}_{n}^{\pm}\mathbf{p}_{n}^{\pm}\right]\frac{i}{2\beta_{n}}
e^{i\beta_{n}\left|z-z^{\prime}\right|}-\frac{1}{k_{n}^{2}}\hat{e}_{z}\hat{e}_{z}\delta\left(z-z^{\prime}
\right),\label{eq:G0_in_Weyl_representation}
\end{equation}
where we have introduced $\mathbf{p}_{n}^{\pm}=\mathbf{p}_{\parallel}\pm\beta_{n}\hat{e}_{z}$,
with the $\pm$ sign applying for $z\gtrless z^{\prime}$. The last term
in the above equation is the depolarization term, that arises from
the boundary contributions when performing integration by parts, due
to the exclusion of an infinitesimal line element around $z^{\prime}=z$
in $\fint$. The principal value in the first term indicates that
a small region around $z^{\prime}=z$ is to be excluded. We also notice,
that this depolarization term could also have been obtained from the
general depolarization dyadic in real space, Eq.~(\ref{eq:depolarization_dyadic}),
if we choose as excluded volume an infinite slab located at $-\delta<z<\delta$
(with $\delta\rightarrow0$). For this excluded volume, we would obtain
$\overleftrightarrow{L}_{\text{Slab}_{\delta}}=\hat{e}_{z}\hat{e}_{z}$. 

It is possible to write Eq.~(\ref{eq:G0_in_Weyl_representation})
in a more meaningful way by introducing the $s$- and $p$-polarization
vectors. The $s$--polarization vector lies in the $xy-$plane and
is therefore written as \cite{anti-critical-angle-A} 
\begin{equation}
\hat{e}_{s}=\frac{\mathbf{p}_{\parallel}\times\hat{e}_{z}}{p_{\parallel}}.
\end{equation}
On the other hand, the $p$--polarization vector is orthogonal to $\mathbf{p}_{n}^{\pm}$
and $\hat{e}_{s}$, and therefore we write it as \cite{anti-critical-angle-A}
\begin{equation}
\hat{e}_{p,n}^{\pm}=\frac{\hat{e}_{s}\times\mathbf{p}_{n}^{\pm}}{k_{n}}=\frac{p_{\parallel}}{k_{n}}\hat{e}_{z}
\mp\frac{\beta_{n}}{k_{n}}\frac{\mathbf{p}_{\parallel}}{p_{\parallel}},
\end{equation}
where $\hat{e}_{p,n}^{\pm}$ is the $p$-polarization vector for a
field propagating in the positive/negative $z-$direction. With these
definitions one obtains the following identity
\begin{equation}
\overleftrightarrow{I}-\frac{1}{k_{n}^{2}}\mathbf{p}_{n}^{\pm}\mathbf{p}_{n}^{\pm}=\hat{e}_{s}\hat{e}_{s}+\hat
{e}_{p,n}^{\pm}\hat{e}_{p,n}^{\pm}.
\end{equation}
Therefore, we can write Eq.~(\ref{eq:G0_in_Weyl_representation})
as \cite{A-note-on-Green,Arnoldus-2015}: 
\begin{equation}
\overleftrightarrow{G_0}\left(\mathbf{p}_{\parallel},z,z^{\prime},\omega\right)=\hat{e}_{s}\hat{e}_{s}\frac{
i}{2\beta_{n}}e^{i\beta_{n}\left|z-z^{\prime}\right|}+\hat{e}_{p,n}^{\pm}\hat{e}_{p,n}^{\pm}\frac{i}{2\beta_{n
}}e^{i\beta_{n}\left|z-z^{\prime}\right|} -\frac{1}{k_{n}^{2}}\hat{e}_{z}\hat{e}_{z}\delta\left(z-z^{\prime}
\right),
\end{equation}
with the first and the second terms corresponding to the $s$-- and $p$-polarization
components of the free-space dyadic Green's function, respectively. A different
derivation of previous two equations has been given in the literature
before \cite{Bedeaux-1973,Sipe-1979,Sipe-1987}.
The same decomposition has been used in the study of an emitter's life-time near a graphene sheet \cite{Nikitin:2011aa,Nikitin:2013aa} and in the context of
the calculation of the electric field of a dipole near graphene \cite{Hanson:2008aa}. 

\subsection{Source and scattered Green's functions: scattering at a planar 
interface}\label{subsec:Source-and-scattered}

We now want to address the problem of determining the Green's function
in a system with a planar interface between two media 1 and 2. To that end, we shall evaluate the electric 
field generated by a
point dipole, characterized by an electric dipole moment $\mathbf{d}_{0}$, located at a
distance $z_{0}>0$ from the interface. We assume that medium 1 is
located in the half-space $z>0$, whereas medium 2 is located in the
complementary space, as represented in Fig.~\ref{fig:System}. Note that in general $\beta_{1}\neq\beta_{2}$
due to the different values of the speed of light in the media.
The field emitted by the oscillating dipole in the half-space $z>0$
reads (assuming that $z\ne z_{0}$) 
\begin{equation}
\mathbf{E}_{0}\left(\mathbf{p}_{\parallel},z,z_{0},\omega\right)=\mu_{1}\mu_{0}\omega^{2}\overleftrightarrow{G
}_{0}\left(\mathbf{p}_{\parallel},z,z_{0},\omega\right)\cdot\mathbf{d}_{0}.\label{eq:emanated_field}
\end{equation}
We have two different values for the field, depending on whether $z\gtrless z_{0}$.
Concretely, we obtain
\begin{equation}
\mathbf{E}_{0}\left(\mathbf{p}_{\parallel},z\gtrless 
z_{0},\omega\right)=\mu_{1}\mu_{0}\omega^{2}\frac{i}{2\beta_{1}}e^{i\beta_{1}\left|z-z_{0}\right|}\left[\hat{e
}_{s}\left(\hat{e}_{s}\cdot\mathbf{d}_{0}\right)+\hat{e}_{p,1}^{\pm}\left(\hat{e}_{p,1}^{\pm}\cdot\mathbf{d}_{
0}\right)\right],
\end{equation}
which we can write as 
\begin{equation}
\mathbf{E}_{0}\left(\mathbf{p}_{\parallel},z\gtrless 
z_{0},\omega\right)=E_{0,s}e^{i\beta_{1}\left|z-z_{0}\right|}\hat{e}_{s}+E_{0,p}e^{i\beta_{1}\left|z-z_{0}
\right|}\hat{e}_{p,1}^{\pm},
\end{equation}
with $s-$ and $p$-polarization amplitudes being given by 
\begin{align}
E_{0,s} & =\mu_{1}\mu_{0}\omega^{2}\frac{i}{2\beta_{1}}\left(\hat{e}_{s}\cdot\mathbf{d}_{0}\right),\\
E_{0,p} & =\mu_{1}\mu_{0}\omega^{2}\frac{i}{2\beta_{1}}\left(\hat{e}_{p,n}^{\pm}\cdot\mathbf{d}_{0}\right).
\end{align}
This field will imping on the interface at $z=0$, being partially
reflected and partially transmitted. The reflected and transmitted
fields can be expressed in terms of the amplitudes of the imping field
at $z=0$ field and of the reflection, $r_{s}$ and $r_{p}$, and
transmission, $t_{s}$ and $t_{p}$, coefficients of the interface
for the $s-$ and $p-$polarizations as \cite{anti-critical-angle-A,Arnoldus-2015}
\begin{align}
\mathbf{E}_{r}\left(\mathbf{p}_{\parallel},z>0,\omega\right) & 
=r_{s}E_{0,s}e^{i\beta_{1}\left(z+z_{0}\right)}\hat{e}_{s}+r_{p}E_{0,p}e^{i\beta_{1}\left(z+z_{0}\right)}\hat{
e}_{p,1}^{+},\\
\mathbf{E}_{t}\left(\mathbf{p}_{\parallel},z<0,\omega\right) & 
=t_{s}E_{0,s}e^{i\beta_{1}z_{0}}e^{-i\beta_{2}z}\hat{e}_{s}+t_{s}E_{0,s}e^{i\beta_{1}z_{0}}e^{-i\beta_{2}z}
\hat{e}_{p,2}^{-}.
\end{align}
The factor $e^{i\beta_{1}z}\left(e^{-i\beta_{2}z}\right)$ is acquired
by the field while propagating along the positive(negative) $z$ direction
in medium 1(2). The $p$-polarization vector for the reflected field
is $\hat{e}_{p,1}^{+}$ since it propagates along the positive $z$ direction. Conversely, 
we have $\hat{e}_{p,2}^{-}$ for the transmitted field, since
it propagates along the negative $z$ direction. 

Therefore, the total field for $z>0$ can be written as 
\begin{equation}
\mathbf{E}\left(\mathbf{p}_{\parallel},z>0,z_{0},\omega\right)=\mu_{1}\mu_{0}\omega^{2}\left[
\overleftrightarrow{G_0}(\mathbf{p}_{\parallel},z-z_{0},\omega)+\overleftrightarrow{G_r}(\mathbf{p}_{
\parallel},z,z_{0},\omega)\right]\cdot\mathbf{d}_{0},
\end{equation}
where we have introduced the reflected Green's function
\begin{equation}
\overleftrightarrow{G_r}\left(\mathbf{p}_{\parallel},z,z_{0},\omega\right)=r_{s}\frac{i}{2\beta_{1}}\hat{e}_
{s}\hat{e}_{s}e^{i\beta_{1}(z+z_{0})}+r_{p}\frac{i}{2\beta_{1}}\hat{e}_{p,1}^{+}\hat{e}_{p,1}^{-}e^{i\beta_{1}
(z+z_{0})}.
\end{equation}
Similarly, the transmitted field for $z<0$ can be written as
\begin{equation}
\mathbf{E}(\mathbf{p}_{\parallel},z<0,z_{0},\omega)=\mu_{1}\mu_{0}\omega^{2}\overleftrightarrow{G_t}
\left(\mathbf{p}_{\parallel},z,z_{0},\omega\right)\cdot\mathbf{d}_{0}.
\end{equation}
with the transmitted Green's function being written as
\begin{equation}
\overleftrightarrow{G_t}\left(\mathbf{p}_{\parallel},z,z_{0},\omega\right)=t_{s}\frac{i}{2\beta_{1}}\hat{e}_
{s}\hat{e}_{s}e^{-i\beta_{2}z}e^{i\beta_{1}z_{0}}+t_{p}\frac{i}{2\beta_{1}}\hat{e}_{p,2}^{-}\hat{e}_{p,1}^{-}
e^{-i\beta_{2}z}e^{i\beta_{1}z_{0}}.
\end{equation}
At this point, we have now in our possession all the relevant tools to study the renormalization of the 
polarizability of a nanoparticle in the vicinity of a planar interface.

\section{Renormalization of the polarizability of a quantum emitter near a
graphene sheet and a graphene-based grating}\label{sec:Renormalization-polarizability}

The dyadic Green's function method is a powerful tool  for describing the modification of the properties of a quantum
emitter near interfaces, as it takes into account the change in the density of electromagnetic modes induced 
by the presence of the interface. Problems such as the calculation of the Purcell factor and F\"orster energy transfer 
are two examples  \cite{Forster-Graphene,Kamp_2015} well suited for the Green's function approach. Here we consider another problem that also depends on the density of electromagnetic modes, the 
calculation of the effective polarizability of a quantum emitter.

\subsection{Polarizability of a quantum emitter in a homogeneous medium}\label{subsec:def-polarizability}

The polarizability of a nanoparticle, $\overleftrightarrow{\alpha}$,
treated as a point objective, relates the electric dipole moment, $\mathbf{d}$,
that is induced in the nanoparticle to the value of the externally
applied electric field, $\mathbf{E}_{\text{ext}}\left(\mathbf{r}_{0}\right)$,
at the nanoparticle's position, $\mathbf{r}_{0}$, via
\begin{equation}
\mathbf{d}=\overleftrightarrow{\alpha}\cdot\mathbf{E}_{\text{ext}}\left(\mathbf{r}_{0}\right).
\end{equation}
Note that $\mathbf{E}_{\text{ext}}\left(\mathbf{r}_{0}\right)$ does
not include self-field effects, that is, the electric field 
generated by the nanoparticle itself when subjected to $\mathbf{E}_{\text{ext}}\left(\mathbf{r}_{0}\right)$.
Let us consider a homogeneous medium characterized by $\epsilon_{1}$ and $\mu_{1}$, 
in which a nanoparticle with dielectric function $\epsilon_{\text{np}}(\omega)$ lives. 
Then,  the electric field obeys Eq.~(\ref{eq:wave_eq_Electric}) with
the free current due to the nanoparticle polarization  (excluding the current to the polarization density of 
the homoegenous medium) being written as
\begin{equation}
\mathbf{j}_{f}(\mathbf{r},\omega)=-i\omega\left[\mathbf{P}_{\text{np}}(\mathbf{r},\omega)-\mathbf{P}_{1}
(\omega)\right]=-i\omega\epsilon_{0}\left(\epsilon_{\text{np}}(\omega)-\epsilon_{1}\right)\mathbf{E}(\mathbf{r
},\omega),
\label{eq:jf_diff_pol}
\end{equation}
where we have used the usual linear consititutive relation 
$\mathbf{P}_{n}(\omega,\mathbf{r})=\epsilon_{0}\left(\epsilon_{1}-1\right)\mathbf{E}(\mathbf{r},\omega)$,
$\mathbf{P}_{\text{np}}(\mathbf{r},\omega)$ is the polarization density
due to the nanoparticle, $\mathbf{P}_{1}(\omega)$ is the polarization
density due to the homogeneous medium, and $\mathbf{E}(\mathbf{r})$
is the total electric field in the nanoparticle.
Therefore, from Eq.~(\ref{eq:electric_field}), the electric field
obeys a Lippmann-Schwinger equation \cite{Lippmann} 
\begin{equation}
\mathbf{E}(\mathbf{r},\omega)=\mathbf{E}_{\text{ext}}(\mathbf{r},\omega)+\omega^{2}\mu_{1}\mu_{0}\int_{V}
d\mathbf{r}^{\prime}\left(\epsilon_{\text{np}}(\omega)-\epsilon_{1}\right)\overleftrightarrow{G_0}(\mathbf{r
},
\mathbf{r}^{\prime},\omega)\cdot\mathbf{E}(\mathbf{r},\omega)\label{eq:nanoparticle_self_consistent}
\end{equation}
where $\mathbf{E}_{\text{ext}}(\mathbf{r},\omega)$ is a solution
of the wave equation in the homogeneous medium, and $V$ is the volume
of the nanoparticle. We want to solve for the electric field inside
the nanoparticle. We will follow the approximate approach of Ref.~\cite{Dahan-2012}.
We shall assume a spherical nanoparticle, with radius $R$, and assume
that $k_{n}R\ll1$. This allows us to approximate $\mathbf{E}(\mathbf{r},\omega)$
as constant inside the nanoparticle and to take the limit 
$\left|\mathbf{r}-\mathbf{r}^{\prime}\right|\rightarrow0$
for $\overleftrightarrow{G_0}(\mathbf{r}-\mathbf{r}^{\prime},\omega)$.
Taking into account Eqs.~(\ref{eq:FF_GF})-(\ref{eq:NF_GF}), we
can write the regular part (excluding the Dirac $\delta$-function) of the free-space dyadic Green's function as
\begin{equation}
\overleftrightarrow{G_0}^{{\rm 
reg}}\left(\mathbf{r},\mathbf{r}^{\prime},\omega\right)=\left(\frac{1}{\left|\mathbf{R}\right|}+\frac{i}{k_{1}
\left|\mathbf{R}\right|^{2}}-\frac{1}{k_{1}^{2}\left|\mathbf{R}\right|^{3}}\right)\overleftrightarrow{I}\frac{
e^{ik_{1}\left|\mathbf{R}\right|}}{4\pi}+\left(-\frac{1}{\left|\mathbf{R}\right|}-\frac{3i}{k_{1}\left|\mathbf
{R}\right|^{2}}+\frac{3}{k_{1}^{2}\left|\mathbf{R}\right|^{3}}\right)\hat{\mathbf{R}}\mathbf{\hat{R}}\frac{e^{
ik_{1}\left|\mathbf{R}\right|}}{4\pi},
\end{equation}
where $\mathbf{R}=\mathbf{r}-\mathbf{r}^{\prime}$ and $\hat{\mathbf{R}}=\mathbf{R}/|\mathbf{R}|$. Performing an
angular average and taking the limit $\left|\mathbf{r}-\mathbf{r}^{\prime}\right|\rightarrow0$,
we obtain
\begin{equation}
\overleftrightarrow{G_0}^{{\rm 
reg}}\left(\mathbf{r},\mathbf{r}^{\prime},\omega\right)\simeq\frac{1}{6\pi}\frac{1}{\left|\mathbf{R}\right|}
\overleftrightarrow{I}+i\frac{k_{1}}{6\pi}\overleftrightarrow{I}.\label{eq:GF_reg_small}
\end{equation}
We neglect the real part of $\overleftrightarrow{G_0}^{{\rm reg}}(\mathbf{r},\mathbf{r}^{\prime},\omega)$ 
when
compared to $\overleftrightarrow{G_0}^{{\rm SF}}(\mathbf{r}-\mathbf{r}^{\prime},\omega)$, thereby approximating 
\begin{equation}
\overleftrightarrow{G_0}\left(\mathbf{r},\mathbf{r}^{\prime},\omega\right)\simeq-\frac{1}{3k_{1}^{2}}
\overleftrightarrow{I}\delta(\mathbf{r}-\mathbf{r}^{\prime})+i\frac{k_{1}}{6\pi}\overleftrightarrow{I}.
\end{equation}
Using the above approximation in Eq.~(\ref{eq:nanoparticle_self_consistent})
and assuming that the electric field within the nanoparticle varies slowly, that is, $\mathbf{E}(\mathbf{r},\omega)=\mathbf{E}(\mathbf{r}_{0},\omega)$ throughout $V$, we obtain
\begin{equation}
\mathbf{E}(\mathbf{r}_{0},\omega)=\mathbf{E}_{\text{ext}}(\mathbf{r}_{0},\omega)+\omega^{2}\mu_{1}\mu_{0}
\left(\epsilon_{\text{np}}(\omega)-\epsilon_{1}\right)\left(-\frac{1}{3k_{1}^{2}}+i\frac{k_{1}}{6\pi}
V\right)\mathbf{
E}(\mathbf{r}_{0},\omega).
\end{equation}
Solving for $\mathbf{E}(\mathbf{r}_{0},\omega)$ we obtain
a relation between the externally applied and the local electric fields
\begin{equation}
\mathbf{E}(\mathbf{r}_{0},\omega)=\frac{1}{1-\frac{1}{\epsilon_{1}}\left(\epsilon_{\text{np}}
(\omega)-\epsilon_{1}
\right)\left(-\frac{1}{3}+i\frac{k_{1}^{3}}{6\pi}V\right)}\mathbf{E}_{\text{ext}}(\mathbf{r}_{0},\omega).
\end{equation}
Therefore, the electric dipole moment follows from
\begin{align}
\mathbf{d} & 
=\epsilon_{0}\left(\epsilon_{\text{np}}(\omega)-\epsilon_{1}\right)\int_{V}d^{3}\mathbf{r}\mathbf{E}(\mathbf{r
},
\omega)\nonumber \\
 & 
\simeq\frac{\alpha_{\rm CM}}{1-i\frac{k_{1}^{3}}{6\pi\epsilon_{0}\epsilon_{1}}\alpha_{\rm CM}}\mathbf{E}_{\text{ext}}
(\mathbf { r
}_{0},\omega),
\label{eq:d_E_0_relation}
\end{align}
where we have introduced the Clausius-Mossotti polarizability 
 \cite{polarizability_simple_bodies} 
\begin{equation}
\alpha_{\rm CM}=4\pi\epsilon_{1}\epsilon_{0}R^{3}\frac{\epsilon_{\text{np}}(\omega)-\epsilon_{1}}{\epsilon_{\text{
np}}
(\omega)+2\epsilon_{1}}.
\end{equation}
The polarizability of a nanoparticle embedded in a homogeneous medium with relative permittivity $\epsilon_1$ can be read from Eq.~(\ref{eq:d_E_0_relation})
\begin{equation}
\overleftrightarrow{\alpha_0}=\frac{\alpha_{\rm CM}}{1-i\frac{k_{1}^{3}}{6\pi\epsilon_{0}\epsilon_{1}}\alpha_{CM
} }
\overleftrightarrow{I}.\label{eq:vacuum_polarizability}
\end{equation}
It must be pointed out that the above equation is only approximate.
As a matter of fact it is easy to see that if we had kept the term
$\propto1/\left|\mathbf{R}\right|$ in $\overleftrightarrow{G_0}^{{\rm 
reg}}(\mathbf{r}-\mathbf{r}^{\prime},\omega)$
we would generated a real term $\propto k_{1}^{2}$ in the denominator
of Eq.~(\ref{eq:vacuum_polarizability}), which can be interpreted
as a dynamic depolarization effect \cite{Meier_1983}. The obtained
term would still be incorrect, as additional terms of the same order
in $k_{1}$ would appear from taking into account that the electric
field inside the nanoparticle is not constant. An exact treatment
using Mie's scattering theory for a spherical particle would lead
to \cite{SERS,polarizability_simple_bodies}
\begin{equation}
\overleftrightarrow{\alpha}_{\text{Mie}}=
4\pi\epsilon_0\epsilon_1R^3
\left[
\frac{\epsilon_{\text{np}}
	(\omega)+2\epsilon_{
		1}}{\epsilon_{\text{np}}(\omega)-\epsilon_{1}}
-\frac{3}{5}\frac{\epsilon_{\text{np}}
(\omega)-2\epsilon_{
1}}{\epsilon_{\text{np}}(\omega)-\epsilon_{1}}R^2k_{1}^{2}-i\frac
{2}{3}
R^3k_{1}^{3}\right]^{-1}\overleftrightarrow{I}.
\end{equation}
There is indeed a term of order $k_{1}^{2}$ but the term of order
$k_{1}^{3}$ is unchanged. The imaginary term of order $k_{1}^{3}$
is usually denoted by radiation damping correction \cite{Novotny,radiation-damping}
and is essentially to enforce the optical theorem for electromagnetic
scattering to lowest order \cite{Novotny,Draine_1994,Stauber_2014}. Notice
that the radiation damping correction is also responsible for the
decay rate of the dipole. As a matter of fact, the transition rate
of a quantum emitter in a homogeneous medium 1 is quantum mechanically
given by (see derivation in Appendix \ref{sec:Derivation-rate})

\begin{equation}
\frac{1}{\tau_{1}}=\frac{2\omega^{2}}{\hbar}\mu_{1}\mu_{0}\mathbf{d}_{0}^{\dagger}\cdot\Im\overleftrightarrow{
G}_{0}(\mathbf{r}_{0},\mathbf{r}_{0},\omega)\cdot\mathbf{d}_{0},\label{eq:rate}
\end{equation}
for a real-valued dipole moment. From Eq.~(\ref{eq:GF_reg_small}), we have
\begin{equation}
\Im\ \overleftrightarrow{G_0}(\mathbf{r}_{0},\mathbf{r}_{0},\omega)=\frac{k_{1}}{6\pi}\overleftrightarrow{I},
\end{equation}
such that

\begin{equation}
\frac{1}{\tau_{1}}=\frac{\omega^{3}}{3\pi 
v_{1}\hbar}\mu_{1}\mu_{0}\mathbf{d}_{0}^{\dagger}\cdot\mathbf{d}_{0}.
\end{equation}
In the next sections, we will ignore the term of order $k_{1}^{2}$
as it plays no significant role. However, it will become clear that
it is essential to keep the radiation damping correction (arising when the self-field is accounted for).

\subsection{Polarizability of a quantum emitter in proximity to a planar 
interface}\label{subsec:renormalization-interface}

If the nanoparticle is situated in the vicinity of an interface, it is also
possible to write an equation of the 
Lippmann-Schwinger type for the electric field
similar to Eq.~(\ref{eq:nanoparticle_self_consistent}). The only
difference is that in order to take into account the interface, the
free-space dyadic Green's must be replaced by other Green's function which incorporates the reflection from the substrate,  for instance,
$\overleftrightarrow{G_0}(\mathbf{r},\mathbf{r}^{\prime},\omega)\rightarrow\overleftrightarrow{G}(\mathbf{r}
,\mathbf{r}^{\prime},\omega)=\overleftrightarrow{G_0}(\mathbf{r},\mathbf{r}^{\prime},
\omega)+\overleftrightarrow{G_r}(\mathbf{r},\mathbf{r}^{\prime},\omega)$.
Likewise, the external field $\mathbf{E}_{0}(\mathbf{r},\omega)$
must be replaced by a solution of the electric field wave equation
in the presence of the substrate,
$\mathbf{E}_{0}(\mathbf{r},\omega)\rightarrow\mathbf{E}_{\text{ext}}(\mathbf{r},\omega)=\mathbf{E}_{0}(\mathbf
{r},\omega)+\mathbf{E}_{r}(\mathbf{r},\omega)$,
where $\mathbf{E}_{0}(\mathbf{r},\omega)$ is the incident field and
$\mathbf{E}_{r}(\mathbf{r},\omega)$ is the reflected field.
Therefore, the Lippmann-Schwinger equation for the electric field
taking into account the substrate is given by
\begin{equation}
\mathbf{E}(\mathbf{r},\omega)=\mathbf{E}_{\text{ext}}(\mathbf{r},\omega)+\omega^{2}\mu_{1}\mu_{0}\int_{V}
d\mathbf{r}^{\prime}\left(\epsilon_{\text{np}}(\omega)-\epsilon_{1}\right)\overleftrightarrow{G}(\mathbf{r},
\mathbf{r
}^{\prime},\omega)\cdot\mathbf{E}(\mathbf{r},\omega).\label{eq:Lippmann}
\end{equation}
We now proceed in the same fashion as before, assuming $k_{1}R\ll1$,
approximating $\mathbf{E}(\mathbf{r},\omega)=\mathbf{E}(\mathbf{r}_{0},\omega)$
as constant inside the nanoparticle, and keeping only the dominant contributions
from $\overleftrightarrow{G}(\mathbf{r},\mathbf{r}^{\prime},\omega)$
in the limit $\vert\mathbf{r}-\mathbf{r}^{\prime}\vert\rightarrow0$.
Therefore we write \cite{Dahan-2012}
\begin{equation}
\overleftrightarrow{G}(\mathbf{r},\mathbf{r}^{\prime},\omega)\simeq-\frac{1}{3k_{1}^{2}}\overleftrightarrow{I}
\delta(\mathbf{r}-\mathbf{r}^{\prime})+i\frac{k_{1}}{6\pi}\overleftrightarrow{I}+\overleftrightarrow{G_r}
(\mathbf{r}_{0},\mathbf{r}_{0},\omega),\label{eq:Green_approx}
\end{equation}
where we have used the fact that $\overleftrightarrow{G_r}(\mathbf{r}_{0},\mathbf{r}_{0},\omega)$
is regular. Introducing Eq. (\ref{eq:Green_approx}) into Eq. (\ref{eq:Lippmann}),
we obtain for the field at the point $\mathbf{r}_{0}$ the result

\begin{equation}
\mathbf{E}(\mathbf{r}_{0},\omega)=\mathbf{E}_{{\rm 
ext}}(\mathbf{r}_{0},\omega)+\frac{k_{1}^{2}}{\epsilon_{1}}\left(\epsilon_{\text{np}}(\omega)-\epsilon_{1}
\right)\left[-\frac{\overleftrightarrow{I}}{3k_{1}^{2}}+V\left(\frac{ik_{1}}{6\pi}\overleftrightarrow{I}
+\overleftrightarrow{G_r}(\mathbf{r}_{0},\mathbf{r}_{0},\omega)\right)\right]\cdot\mathbf{E}(\mathbf{r}_{0},
\omega).
\end{equation}
Solving the previous equation for $\mathbf{E}(\mathbf{r}_{0},\omega)$
leads to

\begin{align}
\mathbf{E}(\mathbf{r}_{0},\omega) & 
=\frac{3\epsilon_{1}}{\epsilon_{\text{np}}(\omega)+2\epsilon_{1}}\left[\overleftrightarrow{I}-\alpha_{\rm CM}
\omega^{2}
\mu_{1}\mu_{0}\left(\frac{ik_{1}}{6\pi}\overleftrightarrow{I}+\overleftrightarrow{G_r}(\mathbf{r}_{0},
\mathbf{r}_{0},\omega)\right)\right]^{-1}\cdot\mathbf{E}_{{\rm ext}}(\mathbf{r}_{0},\omega).
\end{align}
The electric dipole moment is thus given by

\begin{equation}
\mathbf{d}=V\epsilon_{0}\left(\epsilon_{\text{np}}(\omega)-\epsilon_{1}\right)\mathbf{E}(\mathbf{r}_{0},
\omega)=\alpha_{\rm CM}\left[\overleftrightarrow{I}-\alpha_{\rm CM}\omega^{2}\mu\mu_{0}\left(\frac{ik_{1}}{6\pi}
\overleftrightarrow{I}+\overleftrightarrow{G_r}(\mathbf{r}_{0},\mathbf{r}_{0},\omega)\right)\right]^{-1}
\cdot\mathbf{E}_{{\rm ext}}(\mathbf{r}_{0},\omega),
\end{equation}
from which we can identify the effective polarizability 
\begin{align}
\overleftrightarrow{\alpha_{\text{eff}}} & 
=\alpha_{\rm CM}\left[\overleftrightarrow{I}-\mu_{1}\mu_{0}\omega^{2}\alpha_{\rm CM}\left(i\Im\overleftrightarrow{G_0}(\mathbf{r}_{0},\mathbf{r}_{0},\omega)+\overleftrightarrow{G_r}(\mathbf{r}_{0},\mathbf{r}_{0},
\omega)\right)^{-1}\right]^{-1}.\label{eq:renormalized_alpha}
\end{align}
This equation can be expressed in terms of the free-space polarizability 
Eq.~(\ref{eq:vacuum_polarizability}) as
\begin{equation}
\overleftrightarrow{\alpha_{\text{eff}}}=\overleftrightarrow{\alpha_0}\left[\overleftrightarrow{I}-\mu_{1}
\mu_{0}\omega^{2}\overleftrightarrow{G_r}(\mathbf{r}_{0},\mathbf{r}_{0},\omega)\cdot
\overleftrightarrow{\alpha_0}\right]^{-1}.
\end{equation}
Equation (\ref{eq:renormalized_alpha}) has been derived in the literature
before following a similar argumentation 
\cite{renormalized_polarizability,renormalized-polarizability-2,Dahan-2012}. 

The importance of keeping the free-space radiation damping correction,
$i\Im\overleftrightarrow{G_0}(\mathbf{r}_{0},\mathbf{r}_{0},\omega)$,
will now become clear. According to Poynting's theorem, the power dissipated
by the nanoparticle is given by
\begin{equation}
P_{\mathrm{dis}}=\frac{\omega}{2}\Im\left[\mathbf{E}_{{\rm 
ext}}^{\dagger}(\mathbf{r}_{0},\omega)\cdot\overleftrightarrow{\alpha_{\text{eff}}}\cdot\mathbf{E}_{{\rm 
ext}}(\mathbf{r}_{0},\omega)\right].
\end{equation}
This implies that the imaginary part of the diagonal components of
$\overleftrightarrow{\alpha_{\text{eff}}}$ must be positive, since
the dissipated power must be positive. It is easily checked that
if $a$ and $g$ are complex quantities then
\begin{equation}
\Im\frac{a}{1-ag}=\frac{\Im a+\left|a\right|^{2}\Im g}{\left|1-ag\right|^{2}}.
\end{equation}
If $\Im a>0$, but otherwise arbitrary, the requirement that $\Im\frac{a}{1-ag}>0$
demands that $\Im g>0$. Translating this into the problem of the polarizability,
since we have that $\Im\alpha_{\rm CM}\geq0$, the requirement that $\Im\alpha_{\text{eff}}>0$
demands that 
$\Im\left[\overleftrightarrow{G}(\mathbf{r}_{0},\mathbf{r}_{0},\omega)\right]=\Im\left[i\Im\overleftrightarrow
{G}_{0}(\mathbf{r}_{0},\mathbf{r}_{0},\omega)+\overleftrightarrow{G_r}(\mathbf{r}_{0},\mathbf{r}_{0},
\omega)\right]>0$.
This is true in general, and can be understood either classically
as the fact that  $\Im\left[\overleftrightarrow{G}(\mathbf{r}_{0},\mathbf{r}_{0},\omega)\right]$
gives the total power emitted by a point dipole, or quantum mechanically,
since the diagonal elements or $\Im\left[\overleftrightarrow{G}(\mathbf{r}_{0},\mathbf{r}_{0},\omega)\right]$
correspond to a spectral function (a density of electromagnetic states),
that is always positive. However, in general it is not true that 
$\Im\left[\overleftrightarrow{G_r}(\mathbf{r}_{0},\mathbf{r}_{0},\omega)\right]$,
which happens for example when subradiance of a quantum emitter occurs.
Therefore, the requirement that $\Im\alpha_{\text{eff}}>0$, forces
us to keep the free-space radiation damping correction.

\subsection{Renormalized polarizability of an isotropic quantum emitter near
	a continuous graphene sheet}\label{subsec:Renormalized-graphene-sheet}

In what follows we shall consider the case of an isotropic quantum
emitter in close proximity to a graphene sheet.  In the previous
sections, we  have seen how the effective polarizability of a nanoparticle
depends on the reflected Green's function, $\overleftrightarrow{G_r}(\mathbf{r}_{0},\mathbf{r}_{0},\omega)$,
which can be reconstructed from its angular spectrum representation
as
\begin{equation}
\overleftrightarrow{G_r}(\mathbf{r}_{0},\mathbf{r}_{0},\omega)=\int\frac{d^{2}\mathbf{p}_{\parallel}}{
\left(2\pi\right)^{2}}\overleftrightarrow{G_r}\left(\mathbf{p}_{\parallel},z_{0},z_{0},\omega\right).
\label{eq:Gr_real_space}
\end{equation}
As shown in Section~\ref{subsec:Source-and-scattered} the
reflected Green's function in the angular spectrum representation
can be written in terms of the Fresnel reflection coefficients. For a planar
interface covered by graphene, the reflection coefficients are given
by \cite{J_Optics-2013,book}

\begin{align}
r_{s} & 
=\frac{\beta_{1}-\beta_{2}-\mu_{0}\omega\sigma_{T}(\omega)}{\beta_{1}+\beta_{2}+\mu_{0}\omega\sigma_{T}
(\omega)},\label{eq:rs}\\
r_{p} & 
=\frac{\beta_{1}\epsilon_{2}-\beta_{2}\epsilon_{1}+\beta_{1}\beta_{2}\sigma_{L}/(\epsilon_{0}\omega)}{\beta_{1
}\epsilon_{2}+\beta_{2}\epsilon_{1}+\beta_{1}\beta_{2}\sigma_{L}/(\epsilon_{0}\omega)},
\end{align}
where $\sigma_{T}(\omega)$ and $\sigma_{L}(\omega)$ are the transverse
and longitudinal optical conductivities of graphene. Neglecting nonlocal
effects in the conductivities we have $\sigma_{T}(\omega)=\sigma_{L}(\omega)=\sigma(\omega)$,
which we will model with a Drude-like term~\cite{book,Peres_2007, Stauber_2007}
\begin{equation}
\sigma(\omega)=\frac{e^{2}}{4\hbar}\frac{4}{\pi}\frac{\epsilon_{F}}{\hbar\gamma-i\hbar\omega},
\end{equation}
where $\epsilon_{F}$ is graphene's Fermi energy and $\gamma$ is
the broadening factor. The transmission coefficients    $t_{s}$
and $t_{p}$, are related to the reflection coefficients via~\cite{book}
\begin{align}
t_{s} & =1+r_{s},\\
t_{p} & =\frac{\beta_{1}}{\beta_{2}}\sqrt{\frac{\epsilon_{2}}{\epsilon_{1}}}(1-r_{p}).
\end{align}
After performing the integration over the angular variable in Eq.~(\ref{eq:Gr_real_space}),
we obtain that $\overleftrightarrow{G_r}(\mathbf{r}_{0},\mathbf{r}_{0},\omega)$
is diagonal. Rotational invariance along the $z$ direction imposes
that $G_{r}^{xx}(\mathbf{r}_{0},\mathbf{r}_{0},\omega)=G_{r}^{yy}(\mathbf{r}_{0},\mathbf{r}_{0},\omega)$,
which will differ from $G_{r}^{zz}(\mathbf{r}_{0},\mathbf{r}_{0},\omega)$.
The same will be true for the polarizability of the nanoparticle,
which, using Eq.~(\ref{eq:renormalized_alpha}), we can write as
\begin{align}
\alpha_{\text{eff}}^{xx}= & 
\alpha_{\text{eff}}^{yy}=4\pi\epsilon_{0}\epsilon_{1}R^{3}\frac{\tilde{\alpha_0}}{1-\left(k_{1}R\right)^{3}
\mathcal{G}_{r}^{\parallel}\left(\mathbf{r}_{0},\mathbf{r}_{0},\omega\right)\tilde{\alpha_0}},\\
 & 
\alpha_{\text{eff}}^{zz}=4\pi\epsilon_{0}\epsilon_{1}R^{3}\frac{\tilde{\alpha_0}}{1-\left(k_{1}R\right)^{3}
\mathcal{G}_{r}^{zz}\left(\mathbf{r}_{0},\mathbf{r}_{0},\omega\right)\tilde{\alpha_0}},
\end{align}
where we have defined the dimensionless quantities 
$\tilde{\alpha_0}=\alpha_{0}/\left(4\pi\epsilon_{0}\epsilon_{1}R^{3}\right)$ 
with $\alpha_{0}$ the diagonal element of the nanoparticle polarizability,  Eq.~(\ref{eq:vacuum_polarizability}),
$\mathcal{G}_{r}^{\parallel}\left(\mathbf{r}_{0},\mathbf{r}_{0},\omega\right)=\left(4\pi/k_{1}\right)G_{r}^{xx
}\left(\mathbf{r}_{0},\mathbf{r}_{0},\omega\right)
=\left(4\pi/k_{1}\right)G_{r}^{yy}\left(\mathbf{r}_{0},\mathbf{r}_{0},\omega\right)$
and 
$\mathcal{G}_{r}^{zz}\left(\mathbf{r}_{0},\mathbf{r}_{0},\omega\right)=\left(4\pi/k_{1}\right)G_{r}^{zz}
\left(\mathbf{r}_{0},\mathbf{r}_{0},\omega\right)$.
More explicitly, these quantities can be evaluated from 
\begin{align}
\mathcal{G}_{r}^{\parallel}\left(\mathbf{r}_{0},\mathbf{r}_{0},\omega\right) & =\frac{i}{2}\int_0^\infty 
dse^{i2k_{1}z_{0}\sqrt{1-s^{2}}}s\left(\frac{1}{\sqrt{1-s^{2}}}r_{s}-\sqrt{1-s^{2}}r_{p}\right),\\
\mathcal{G}_{r}^{zz}\left(\mathbf{r}_{0},\mathbf{r}_{0},\omega\right) & =i\int_0^\infty 
dse^{i2k_{1}d\sqrt{1-s^{2}}}\frac{s^{3}}{\sqrt{1-s^{2}}}r_{p},
\end{align}
where $s=p_\parallel/k_1$.

Some insight on the previous expressions can be
obtained by estimating them in the electrostatic limit, valid for
$k_{1}z_{0}\ll1$. In this limit, the main contribution is due to
the $r_{p}$ reflection coefficient. Approximating 
$\sqrt{1-s^{2}}\simeq\sqrt{\epsilon_{2}/\epsilon_{1}-s^{2}}\simeq is$
we obtain
\begin{equation}
\mathcal{G}_{r}^{zz}\left(\mathbf{r}_{0},\mathbf{r}_{0},\omega\right)\simeq2\mathcal{G}_{r}^{\parallel}
\left(\mathbf{r}_{0},\mathbf{r}_{0},\omega\right)\simeq\int_0^\infty 
dse^{-2k_{1}z_{0}s}s^{2}r_{p},\label{eq:G_refl_static}
\end{equation}
with the reflection coefficient being approximated by
\begin{align}
r_{p} & 
=1-\frac{2\beta_{2}\epsilon_{1}}{\beta_{1}\epsilon_{2}+\beta_{2}\epsilon_{1}+\beta_{1}\beta_{2}\sigma_{L}
/(\epsilon_{0}\omega)}\nonumber \\
 & 
\simeq1-\frac{2\epsilon_{1}}{\epsilon_{2}+\epsilon_{1}}\frac{k_{\text{spp}}(\omega)}{k_{\text{spp}}(\omega)-k_
{1}s},\label{eq:r_p_static}
\end{align}
where 
\begin{equation}
k_{\text{spp}}(\omega)=\frac{\omega}{c}\frac{\epsilon_{1}+\epsilon_{2}}{4\alpha_{f}}\frac{
\hbar\omega+i\hbar\gamma}{\epsilon_{F}},
\end{equation}
is graphene's surface plasmon polariton wavenumber (including dissipation
effects) and $\alpha_f \simeq 1/137$ is the fine structure constant. From these results we can already 
estimate when the effect
of the graphene substrate on the NP polarizability will be more significant.
From Eq.~(\ref{eq:r_p_static}), $r_{p}$ is peaked at $s=\Re k_{\text{spp}}(\omega)/k_{1}$,
while the term $e^{-2k_{1}z_{0}s}s^{2}$ in the integrand of Eq.~(\ref{eq:G_refl_static})
has a maximum at $s=\left(k_{1}z_{0}\right){}^{-1}$. Therefore, 
$\mathcal{G}_{r}^{zz}\left(\mathbf{r}_{0},\mathbf{r}_{0},\omega\right)$
will have a maximum, when this two peaks coincide \cite{Kamp_2015}
which occurs for $\Re k_{\text{spp}}(\omega)z_{0}\simeq1$. In the
eletrostatic limit, Eq.~(\ref{eq:G_refl_static}) can be written in
terms of known functions as
\begin{equation}
\mathcal{G}_{r}^{zz}\left(\mathbf{r}_{0},\mathbf{r}_{0},\omega\right)\simeq2\mathcal{G}_{r}^{\parallel}
\left(\mathbf{r}_{0},\mathbf{r}_{0},\omega\right)\simeq\left(\frac{k_{\text{spp}}(\omega)}{k_{1}}\right)^{3}
f\left(2k_{\text{spp}}(\omega)z_{0}\right),\label{eq:G_analytic}
\end{equation}
where the function $f(z)$ is given by 
\begin{equation}
f(z)=\frac{2}{z^{3}}+\frac{2\epsilon_{1}}{\epsilon_{1}+\epsilon_{2}}\left(\frac{1}{z^{2}}+\frac{1}{z}+e^{-z}
\left[i\pi-\text{Ei}(z)\right]\right),
\end{equation}
with $\text{Ei}(z)$ the exponential integral function, which for
real positive argument is written as $\text{Ei}(x)=-\fint_{-x}^{\infty}dte^{-t}/t.$
However, we point out that Eq.~(\ref{eq:G_analytic}) is valid even
in the presence of finite broadening $\gamma$ in graphene.

We shall consider both metallic and polar semiconductor nanoparticles,
with the relative dielectric function described, respectively, by
Drude and Lorentz models. The Drude model for the dielectric function
reads
\begin{equation}
\epsilon_{\text{Drude}}(\omega)=1-\frac{\omega_{p}^{2}}{\omega(\omega+i\hbar\gamma)}
\end{equation}
where $\omega_{p}$ is the metal's plasma frequency and $\gamma$ is the relaxation
rate, while the Lorentz model for the dielectric function of a polar
material is given by
\begin{equation}
\epsilon_{\text{Lorentz}}(\omega)=\epsilon_{\text{\ensuremath{\infty}}}\left(1+\frac{\omega_{{\rm 
L0}}^{2}-\omega_{{\rm T0}}^{2}}{\omega_{{\rm T0}}^{2}-\omega^{2}-i\omega\Gamma_{{\rm TO}}}\right),
\end{equation}
where $\omega_{\text{TO}}$ and $\omega_{\text{LO}}$ are the frequencies
of the transverse and longitudinal optical phonons, $\Gamma_{\text{TO}}$
is a phonon decay rate, and $\epsilon_{\infty}$ is the high frequency
limit of the dielectric function. 
As examples of commonly used materials
for the production of nanoparticles, we consider gold (metallic) and
 CdSe (polar semiconducing) nanoparticles. Typical values of the
polarizability for different substances are give in Ref. \cite{Handbook}.
The used values for the Lorentz model of CdSe are taken from Ref.
\cite{Mikhail-CdSe}. 

\begin{figure}
	\centering
	\includegraphics[scale=0.6]{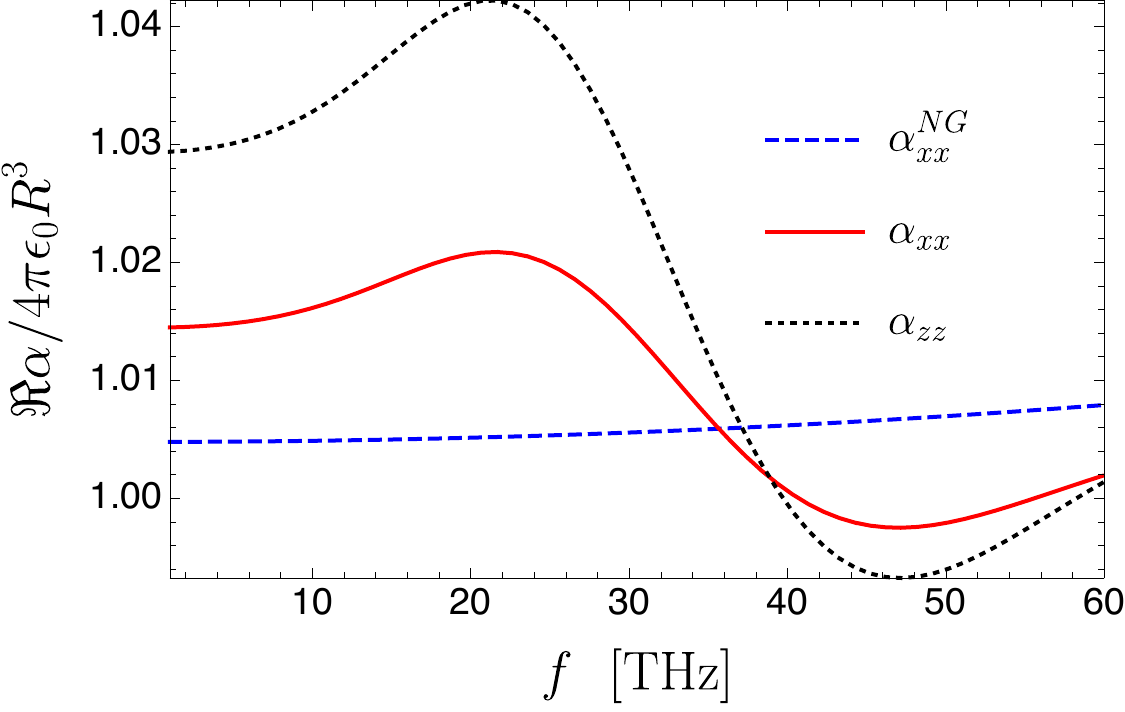}
	\includegraphics[scale=0.6]{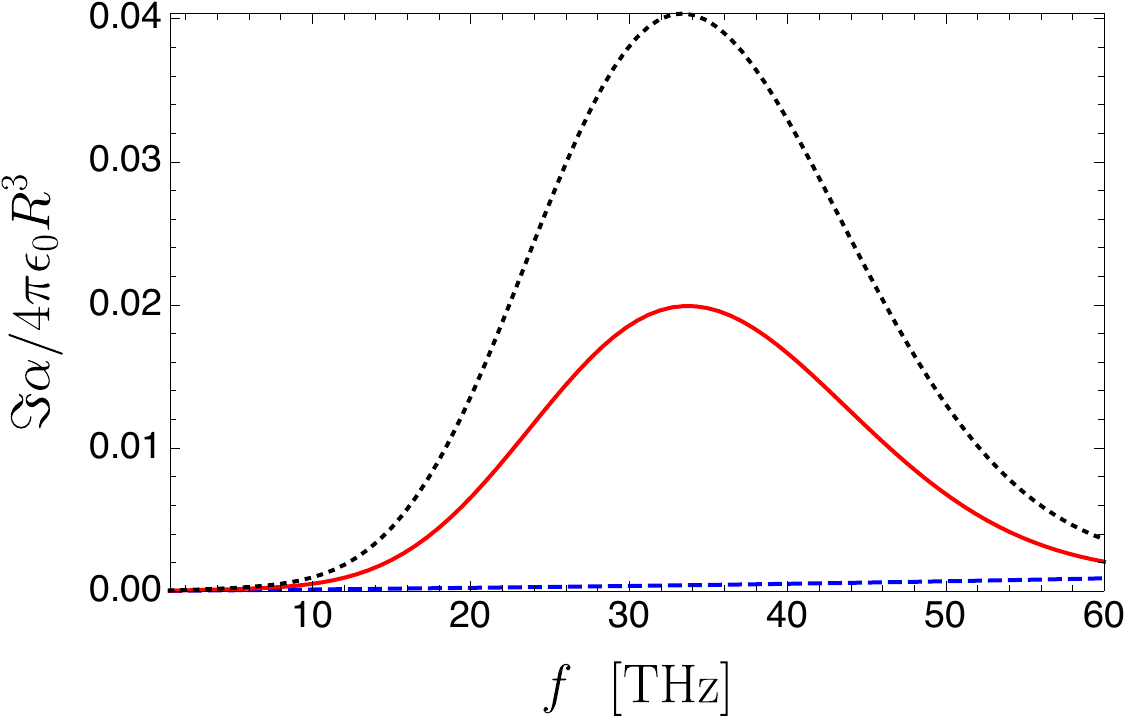}
	\includegraphics[scale=0.6]{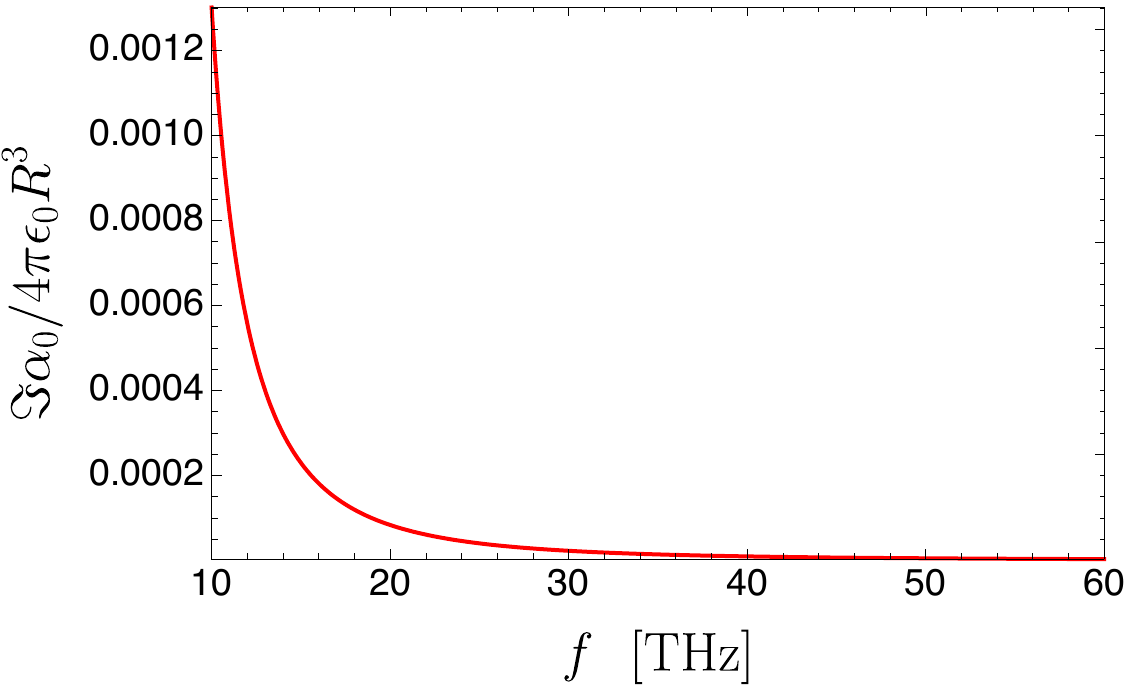}
	\caption{
	Renormalized polarizability of a Gold nanoparticle with $R=50\,$nm
		located at a distance of $z_0=151\,$nm from a graphene sheet
		with a Fermi energy of 1 eV supported by a dielectric of permittivity
		$\epsilon_{2}=2$. The parameters for Gold used in the Drude model
		are: $\omega_{p}=7.9$ eV and $\Gamma_{0}=0.053$ eV. The dashed blue
		line is the polarizability in the absence of graphene (but with the interface present), the solid red
		line represents the component $\alpha_{xx}$, and the black dotted
		line represents the component $\alpha_{zz}$ of the polarizability
		in the presence of graphene.
		The lower panel depicts the polarizability of the nanoparticle in vacuum. One can appreciate the increase in the imaginary part of the polarizability 
		by about two orders of magnitude
		when the particle is near doped graphene.
		\label{fig:Renormalized-polarizability-Drude}}
\end{figure}

In Fig. \ref{fig:Renormalized-polarizability-Drude} we depict the
real and imaginary parts of the polarizability of a Gold nanoparticle
near a doped graphene sheet on a substrate with $\epsilon_{2}=2$.
 The figure clearly shows the strong renormalization of the polarizability of
the nanoparticle relative to its value in the presence of the interface without graphene (blue dashed line).
This is due to the close proximity of the nanoparticle to the graphene sheet, $z_0=151\,$nm.
Nowadays, with the ubiquitous use of hexagonal Boron Nitride (h-BN) for encapsulating graphene, together with 
the possibility of controlling the number of layers of h-BN, it poses no difficulty to routinely produce structures 
where nanoparticles are positioned very close to the graphene sheet, that is, at distances much smaller than their radius.
Also the $zz$--component of the polarizability tensor (black dotted
line) is renormalized differently from the $xx$--component (red
solid line). This is consequence of breaking the translation symmetry along
the $z$--direction introduced by the graphene sheet and the dielectric
change as we cross the $z=0$ plane. We have verified that the broadband 
resonance seen in the imaginary part of the polarizability tensor is due 
to the excitation of surface plasmon-polaritons in graphene. 
This was assessed studying the dispersion of the resonance as a function of the Fermi energy (more on this below).

Given the close proximity of the nanoparticle to the graphene sheet, the question of the  necessity of a nonlocal
description of the graphene conductivity arises. In order to check the correctness of our local description, we have
performed simulations (results not shown) using the nonlocal Drude-like conductivity~\cite{book} of graphene. We have found that 
nonlocality plays no visible role in both the position and the intensity of the resonance in the effective polarizability of the nanoparticle (when $z_0=151$ nm). 
The reasons for this are two-fold: $z_0=151\,$nm of separation between graphene and the nanoparticle is not yet in the range
of $z_0<10\,$nm, where nonlocal effects in metallic 
nanoparticles usually  arise \cite{David:2010aa,Maack:2017aa} (the situation is different for semiconductor nanoparticles
\cite{Maack:2017aa}); the nanoparticle is described as a local dipole and therefore nonlocal effects
play no role in it (only in graphene).

\begin{figure}
	\centering
	\includegraphics[scale=0.6]{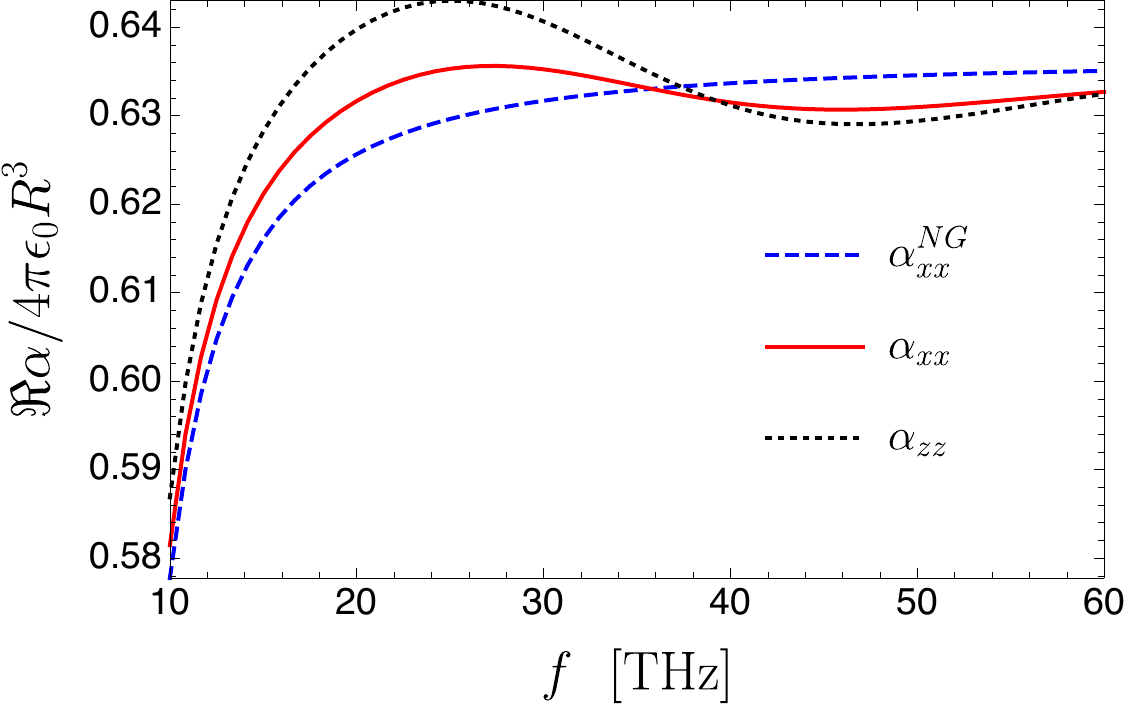}
	\includegraphics[scale=0.6]{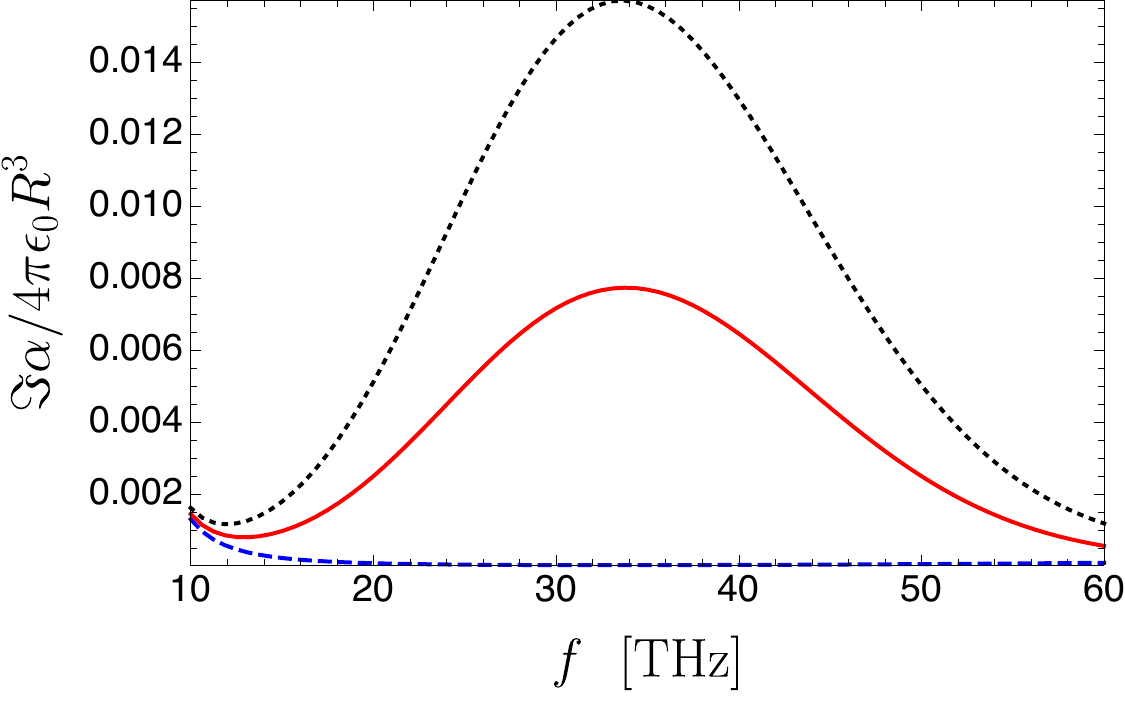}
	\caption{Renormalized polarizability of a CdSe nanoparticle with $R=50\,$nm
		located at a distance of $z_0=151\,$nm from a graphene sheet
		with a Fermi energy of 1 eV supported by a dielectric of permittivity
		$\epsilon_{2}=2$. The parameters used in the Lorentz model are: $\epsilon_{\infty}=6.2$,
		$\omega_{{\rm LO}}=211$ cm$^{-1}$, $\omega_{{\rm TO}}=169$ cm$^{-1}$,
		and $\Gamma_{{\rm TO}}=5$ cm$^{-1}$. The dashed blue line is the
		polarizability in the absence of graphene, the solid red line represents
		the component $\alpha_{xx}$, and the black dotted line represents
		the component $\alpha_{zz}$ of the polarizability in the presence
		of graphene.\label{fig:Renormalized-polarizability-Lorentz}}
\end{figure}

In Fig. \ref{fig:Renormalized-polarizability-Lorentz} we depict
the polarizability of a CdSe nanoparticle in the presence of graphene
on a substrate. As in Fig. \ref{fig:Renormalized-polarizability-Drude},
the  observed broad band resonance in the imaginary part of the polarizability tensor is due to the 
excitation of surface plasmons in graphene. As discussed previously, the order of magnitude of the plasmonic 
resonance frequency can be estimated from $k_{\text{spp}}(\omega)z_0\simeq1$.
When the numbers are pluged in the previous equation, the result is the ball park of the observed resonance in the polarizability spectrum.
In order to further access the plasmonic nature of the broad band resonance, we have studied its position as function of the Fermi energy and found
a complete agreement with the previous equation, that is, the peak of the resonance disperses with $\sqrt{E_F}$.
Interestingly, the intensity of the resonance is smaller by a factor of 3
when compared to the case of the metallic nanoparticle. 
Therefore the latter experiences a strong renormalization of its polarizability in the
presence of a graphene sheet. Note that this will not happen in the presence of a metallic substrate,  for the same studied spectral range, as plasmons in metals
at these frequencies are essentially free radiation.

\subsection{Renormalized polarizability of an isotropic quantum emitter near
	a plasmonic graphene grating}\label{subsec:Renormalized-graphene-grating}

In this section we revisit the problem of the renormalization of the
polarizability of a quantum emitter now considering it near a plasmonic
graphene grating. The used procedure is only approximate, relying
on a semi-analytic approach. However, the analysis performed is sufficient
to capture the effect of plasmonic ressonances of the graphene grating
in the nanoparticle polarizability.

\subsubsection{Optical properties of a plasmonic graphene 
grating}\label{subsubsec:grating-optical}

\begin{figure}
	\centering
	\includegraphics[scale=0.6]{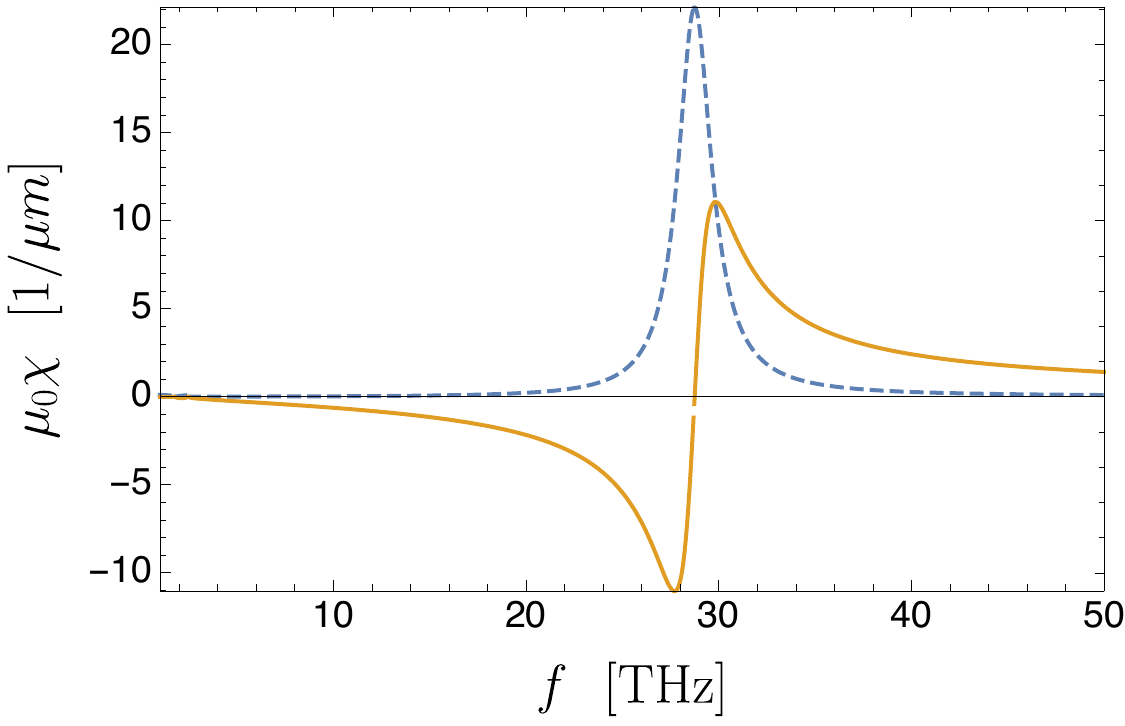}
	\caption{Real (blue dashed line) and imaginary (orange line) of the function
		$\mu_{0}\chi(\omega)$. The parameters of the grating are $L=0.5\,\mu$m
		and $w=L/2$. The Fermi energy of graphene is $E_{F}=1$ eV. The real
		part has a pronounced resonance due to the excitation of a surface
		plasmon-polariton of that frequency ($\sim$87 THz).\label{fig:chi-function}}	
\end{figure}

For metamaterial as the graphene-based grating depicted in 
figure \ref{fig:System} the description of the  interaction of the material with a quantum emitter can be quite complex. 
One possibility to overcome such difficulty is computing
the effective conductivity of the metamaterial, in this case the plasmonic graphene grating. The general method for accomplishing this was given in Ref. \cite{Costas}
and was later applied to the problem of tuning total absorption in
graphene \cite{perfect-absorber}, but no details of its calculation
were given. Instrumental to the calculation of the effective conductivity
is the knowledge of the reflection and transmission Fresnel coefficients.
These were computed in approximated analytical form in Ref. \cite{edge-condition}
and we give here only the final results:

\begin{align}
\label{eq:rpm}
r_{p,m} & =-\delta_{m,0}+t_{p,m}+\mu_0\chi(\omega)\frac{w}{4}J_1(m\pi w/L)\\
t_{p,m} & =\frac{\epsilon_2\beta_{1,m}}{\epsilon_1\beta_{2,m}+\epsilon_2\beta_{1,m}}\left(
2\delta_{m,0}-\mu_0\chi(\omega)\frac{w}{4}J_1(m\pi w/L)
\right)
\end{align}
where $r_{p,0}$ and $t_{p,0}$ are the reflection and transmission
coefficients, respectively, of the zero diffraction-order of the grating
(the only propagating order for a sub-wavelength grating), $w$ is the width of the graphene ribbons
in the grating, $L$ is the period of the grating, and the function
$\chi(\omega)$ reads 
\begin{equation}
\chi(\omega)=\frac{2\beta_{2,0}\beta_{1,0}}{\epsilon_{1}\beta_{2,0}+\epsilon_{2}\beta_{1,0}}\frac{\sigma_{L}(\omega)c^{2}}{\omega\Lambda(\omega)}
\end{equation}
which encodes the information about the plasmonic resonance in the grating, and
with $\Lambda(\omega)$ given by 
\begin{equation}
\Lambda(\omega)=\frac{w}{4}\sum_{n=-\infty}^{\infty}\frac{1}{n}J_{1}(n\pi w/L)\left[1+\frac{\sigma_{L}(\omega)}{\omega\epsilon_{0}}\frac{\beta_{2,n}\beta_{1,n}}{\epsilon_{1}\beta_{2,n}+\epsilon_{2}\beta_{1,n}}\right]
\end{equation}
where $\beta_{1,n}=\sqrt{k_{1}^{2}-k_{x}^{2}-q_{n}^{2}}$ and $\beta_{2,n}=\sqrt{k_{2}^{2}-k_{x}^{2}-q_{n}^{2}}$,
with $q_{n}=k_{y}+n2\pi/L$, $J_{1}(x)$ is the Bessel function of
order 1, and where the summation in $\Lambda(\omega)$ is delicate due to the oscillatory nature of
the Bessel function; see Ref. \cite{edge-condition}. For simplicity of the calculation, we approximate
$\beta_{j,n\ne0}$ by $\beta_{j,n\ne0}\approx\sqrt{k_{j}^{2}-p_{\parallel}^{2}-n^{2}4\pi^{2}/L^{2}}$.
In addition to $r_{p,0}$ and $t_{p,0}$ there is an infinite number
of other coefficients associated with higher diffraction-order, but
they are all evanescent in nature for the parameters chosen in the
figures. Therefore, we approximate the optical properties of the grating
considering only $r_{p,0}$ and $t_{p,0}$, and $r_{p,1}$ and $t_{p,1}$ (we have checked that introducing more evanescent terms does not change the results). This gives us an analytical
description of its optical properties. As noted above, from the knowledge
of $r_{p,0}$ and $t_{p,0}$,  and $r_{p,1}$ and $t_{p,1}$  we can derive an effective conductivity
for the graphene grating along the direction perpendicular to the
axis of the graphene ribbon. This effective conductivity shows a maximum
in its real part associated with the excitation of surface plasmon-polaritons.
The same information is encoded in the function $\chi(\omega)$, as
can be seen in figure \ref{fig:chi-function} and, in fact, for our analysis this latter function is all we need for including 
plasmonic effects into the calculation.

Notice that the conductivity of the system is no longer isotropic. Therefore, we will introduce this anisotropy
in an effective way, choosing different Fermi energies for the $r_s$ and $r_p$ reflection coefficients. 
Also, whereas the $r_{p,m}$ coefficients are given by equation (\ref{eq:rpm}), the $r_s$ coefficient is given by 
equation (\ref{eq:rs}).
This procedure renders our results qualitative and no quantitative agreement is expected  with an exact calculation.
The exact solution would require to extend the formalism to the case on a non-isotropic system in the $xy-$plane. 
Note that this system has broken rotational symmetry around the $z-$axis. Therefore we expected
that the equality $\alpha_{xx}=\alpha_{yy}$ seen in the case of continuous graphene sheet should not hold in the case of grating. Our qualitative results show that this is indeed the case.

\subsubsection{Renormalization of the polarizability of a quantum emitter}

\begin{figure}
	\includegraphics[scale=0.6]{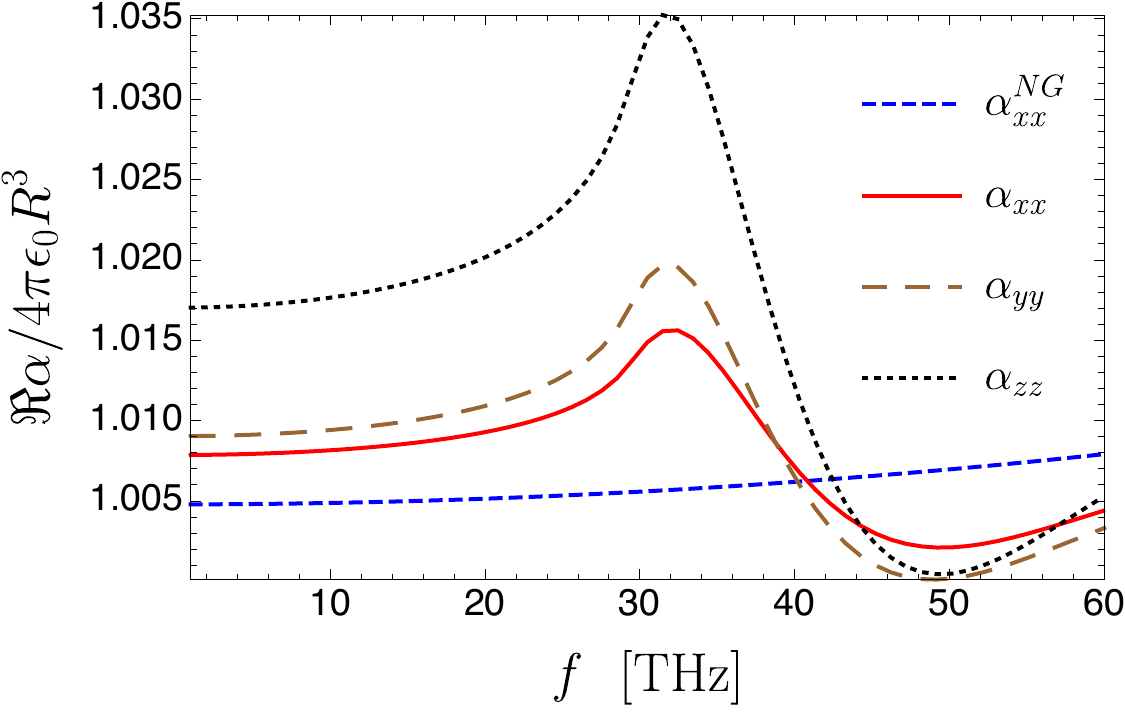}	
	\includegraphics[scale=0.6]{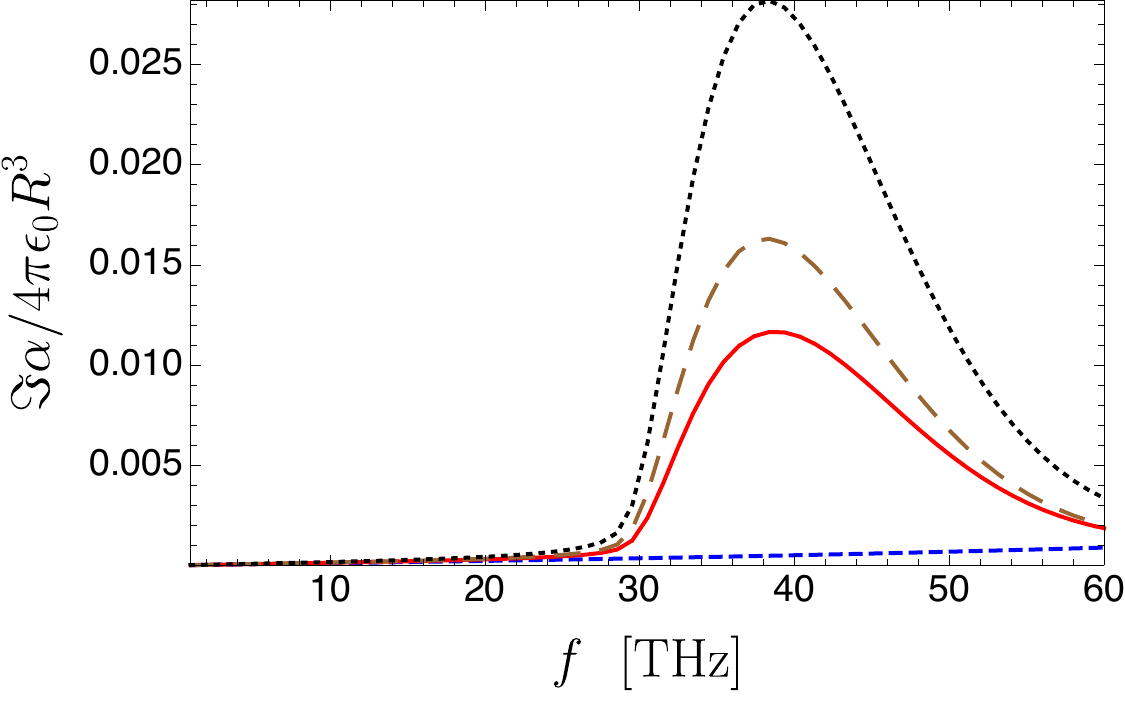}
	\caption{Renormalized polarizability of a Gold nanoparticle in close proximity
		to a plasmonic graphene-based grating. The parameters are the same
		as those in figure \ref{fig:Renormalized-polarizability-Drude}. The
		dashed blue line is the polarizability in the absence of graphene,
		the solid red line represents the component $\alpha_{xx}$,  the dashed brown line represents $\alpha_{yy}$, and
		the black dotted line represents the component $\alpha_{zz}$ of the
		polarizability in the presence of graphene. 
		Note that $\alpha_{xx}\ne\alpha_{yy}$, due to  lack of rotational symmetry in the $xy-$plane introduced by the ribbons structure.
		The parameters of the grating are $L=0.5\,\mu$m
		and $w=L/2$. \label{fig:Renormalized-polarizability-of-grating}}
	
\end{figure}

In this section we study the renormalization of the polarizability
of a quantum emitter near a plasmonic graphene-based grating. As explained
above, we use the reflection coefficients $r_{p,0}$ and $r_{p,1}$ in the reflected
$p$--Green's function and an effective Fermi energy, given by $E_{F}^{{\rm eff}}=E_{F}w/L$
in the $r_{s}$ coefficient, Eq. (\ref{eq:rs}), and use this
in the reflected $s$--Green's function. We consider only the case
of a metallic nanoparticle, as the results are qualitatively the same
for a semiconductor one. In figure \ref{fig:Renormalized-polarizability-of-grating}
we depict the real and imaginary parts of the renormalized polarizability
of a Gold nanoparticle in the proximity of a graphene-based grating.
A strong renormalization of the real part of the 
polarizability  can be seen at the same frequency where the
grating supports the excitation of surface plasmon-polaritons (see
figure \ref{fig:chi-function}). The same happens in the imaginary part.
However, the relative change of the imaginary part is much larger
than for the real part. The results for the imaginary part of the polarizability
in the  case of grating should be compared to those given in figure \ref{fig:Renormalized-polarizability-Drude}
for the same quantity. For the continuous sheet the enhancement of
the imaginary part of $\alpha$ is about twice  the one we have found
in the present case.  This is attributed 
to the approximate description of the reflection coefficients
of the grating. Indeed, we would expect the renormalization to be larger in the  case of the grating
as the latter supports excitation of plasmons by far field radiation, whereas in the 
case of the continuous graphene sheet the excitation of plasmons is due to near-field excitation only.  We also note that the resonance peak in the imaginary part of the polarizability  is not broad-band when compared to the same quantity in the continuous case.

On other hand, the frequency where the maximum of the 
resonance is located  is larger in the present case. This happens since
we can tune the position of the resonance in the grating by varying both
the Fermi energy and the geometric parameters of the grating. Therefore,
the grating system has a versatility that cannot be found in the continuous
sheet case. Indeed using gratings with smaller period, the resonance
can be tuned across the electromagnetic spectrum, from the THz to
the infrared. We also note that the renormalization of the $\alpha_{zz}$
component (black dotted line) is substantially larger than the $\alpha_{xx}$
component (red solid line) and the $\alpha_{yy}$ one (brown dashed line). This happens because the $zz-$component
of the Green's function is about twice as large, compared to the $xx-$component. Finally, we have verified that when $w\rightarrow L$ we recover the results of a continuous graphene sheet.

\section{Extension of the formalism when the quantum emitter has both an electric and a magnetic 
dipole}\label{sec:Inclusion-of-magnetic}

A current density $\mathbf{j}_{f}(\mathbf{r},\omega)$ of a particle
can be described in  terms of its moments in a multipole expansion~\cite{Jackson}. 
A small particle, however, can often be described using only the multipole moments of the lowest orders.
In the case of a metallic nanoparticle, 
its response is dominated by the electric dipole moment. 
Nevertheless, it is
known that in some cases it is necessary to go beyond the electric
dipole approximation and consider higher order moments  
\cite{renormalized-polarizability-2}. In particular,
it has been shown that silicon nanoparticles with size between the
tens and hundreds of nanometers can have a strong responses in the
infrared and visible due to higher order moments~\cite{nl2012_Bozh,scirep12,Mie-Coefficients,magnetic-dipole,renormalized-polarizability-2},  
with the magnetic dipole moment contributing the most, even though the particles are not magnetic by themselves. This
motivate us to generalize the formalism of the previous sections to
the case of a point-like nanoparticle (or quantum emitter) with both electric and magnetic dipole moments. 
Although
the  of Green's functions technique  has been used before in this problem~\cite{Sipe-1987,SNOM-GREEN,renormalized-polarizability-2},
some details regarding the behavior 
of the Green's functions at coincidence, that is, when 
$\mathbf{r}^{\prime}=\mathbf{r}$,  have been overlooked. Therefore,
we carefully present the full formalism, that is, accounting  for both electric and magnetic dipole contributions, below. 

\subsection{Free-space electric, magnetic and mixed Green's functions }

Our starting point are the inhomogeneous Helmholtz equations for the
electric and the magnetic fields (in fact the magnetic field induction $\mathbf{B}(\mathbf{r},\omega)$) in the presence of a source current
density (see Appendix \ref{sec:Derivation-of-wave_equation} for the
derivation):
\begin{align}
-\nabla^{2}\mathbf{E}(\mathbf{r},\omega)-k_{n}^{2}\mathbf{E}(\mathbf{r},\omega) & 
=i\omega\mu_{n}\mu_{0}\left[\mathbf{j}_{f}(\mathbf{r},\omega)+\frac{1}{k_{n}^{2}}
\nabla\left(\nabla\cdot\mathbf{j}_{f}(\mathbf{r},\omega)\right)\right]\label{eq:Helmholtz_E-1}\\
-\nabla^{2}\mathbf{B}(\mathbf{r},\omega)-k_{n}^{2}\mathbf{B}(\mathbf{r},\omega) & 
=\mu_{n}\mu_{0}\nabla\times\mathbf{j}_{f}(\mathbf{r},\omega).\label{eq:Helmholtz_B-1}
\end{align}
As before, we can write the solution for the inhomogeneous Helmholtz
equations as 
\begin{align}
\mathbf{E}(\mathbf{r},\omega) & =\mathbf{E}_{0}(\mathbf{r},\omega)+i\omega\mu_{n}\mu_{0}\int_{\backslash 
V_{\delta}(\mathbf{r})}d^{3}\mathbf{r}^{\prime}g_{0}\left(\mathbf{r},\mathbf{r}^{\prime},
\omega\right)\left(\overleftrightarrow{I}+\frac{1}{k_{n}^{2}}\nabla^{\prime}\nabla^{\prime}\right)\mathbf{j}_{
f}(\mathbf{r}^{\prime},\omega)\label{eq:Helmholtz_E-1-1}\\
\mathbf{B}(\mathbf{r},\omega) & =\mathbf{B}_{0}(\mathbf{r},\omega)+\mu_{n}\mu_{0}\int_{\backslash 
V_{\delta}(\mathbf{r})}d^{3}\mathbf{r}^{\prime}g_{0}\left(\mathbf{r}-\mathbf{r}^{\prime},\omega\right)\nabla^{
\prime}
\times\mathbf{j}_{f}(\mathbf{r}^{\prime},\omega).
\label{eq:Helmholtz_B-1-1}
\end{align}
In the same spirit of Eq. (\ref{eq:jf_diff_pol}), we write the current in terms of a polarization, $\mathbf{P}_{f}$,
and magnetization, $\mathbf{M}_{f}$, densities as
\begin{equation}
\mathbf{j}_{t}(\mathbf{r},\omega)=-i\omega\mathbf{P}_{f}(\mathbf{r},\omega)+\nabla\times\mathbf{M}_{f}(\mathbf
{r},\omega).
\end{equation}
Inserting the latter result  into Eqs.~(\ref{eq:Helmholtz_E-1-1}) and (\ref{eq:Helmholtz_B-1-1})
we obtain
\begin{align}
\mathbf{E}(\mathbf{r},\omega) & =\mathbf{E}_{0}(\mathbf{r},\omega)+\omega^{2}\mu_{n}\mu_{0}\int_{\backslash 
V_{\delta}(\mathbf{r})}d^{3}\mathbf{r}^{\prime}g_{0}\left(\mathbf{r},\mathbf{r}^{\prime},
\omega\right)\left(\overleftrightarrow{I}+\frac{1}{k_{n}^{2}}\nabla^{\prime}\nabla^{\prime}\right)\mathbf{P}_{
f}(\mathbf{r}^{\prime},\omega)\nonumber \\
 & +i\omega\mu_{n}\mu_{0}\int_{\backslash 
V_{\delta}(\mathbf{r})}d^{3}\mathbf{r}^{\prime}g_{0}\left(\mathbf{r},\mathbf{r}^{\prime},\omega\right)\nabla^{
\prime}
\times\mathbf{M}_{f}(\mathbf{r}^{\prime},\omega),\label{eq:Helmholtz_E-1-1-1}\\
\mathbf{B}(\mathbf{r},\omega) & =\mathbf{B}_{0}(\mathbf{r},\omega)-i\omega\mu_{n}\mu_{0}\int_{\backslash 
V_{\delta}(\mathbf{r})}d^{3}\mathbf{r}^{\prime}g_{0}\left(\mathbf{r},\mathbf{r}^{\prime},
\omega\right)\nabla\times\mathbf{
P}_{f}(\mathbf{r}^{\prime},\omega)\nonumber \\
 & +\mu_{n}\mu_{0}\int_{\backslash 
V_{\delta}(\mathbf{r})}d^{3}\mathbf{r}^{\prime}g_{0}\left(\mathbf{r},\mathbf{r}^{\prime},
\omega\right)\left(-\nabla^{
\prime2}+\nabla^{\prime}\nabla^{\prime}\right)\mathbf{M}_{f}(\mathbf{r}^{\prime},\omega),\label{
eq:Helmholtz_B-1-1-1}
\end{align}
where we have use the fact that 
$\nabla^{\prime}\cdot\left(\nabla^{\prime}\times\mathbf{M}_{f}(\mathbf{r}^{\prime},\omega)\right)=0$
and 
$\nabla^{\prime}\times\nabla^{\prime}\times\mathbf{M}_{f}(\mathbf{r}^{\prime},\omega)=\nabla^{\prime}
\left(\nabla^{\prime}\cdot\mathbf{M}_{f}(\mathbf{r}^{\prime},\omega)\right)-\nabla^{\prime2}\mathbf{M}_{f}
(\mathbf{r}^{\prime},\omega)$.
We now proceed as in section~\ref{subsec:Free-space}, using
integration by parts, while taking into account the boundary terms
due to the excluded volume $V_{\delta}$ enclosing the point  $\mathbf{r}^{\prime}=\mathbf{r}$,
in the same form we have already dealt with the electric field Green's function before.
The crossed terms relating the magnetization to the electric field
and the polarization to the magnetic field, only involve one derivative
of the Helmholtz Green's function and therefore the generated boundary
term vanishes in the limit of infinitesimal excluded volume. Therefore,
we may simply write 
\begin{equation}
\int_{\backslash 
V_{\delta}(\mathbf{r})}d^{3}\mathbf{r}^{\prime}g_{0}\left(\mathbf{r},\mathbf{r}^{\prime},\omega\right)\nabla^{
\prime}
\times\mathbf{M}_{f}(\mathbf{r}^{\prime},\omega)=\int_{\backslash 
V_{\delta}(\mathbf{r})}d^{3}\mathbf{r}^{\prime}\nabla 
g_{0}\left(\mathbf{r},\mathbf{r}^{\prime},\omega\right)\times\mathbf{M}_{f}(\mathbf{r}^{\prime},\omega),
\end{equation}
where we have used the fact that in a translation invariant system
$\nabla^{\prime}g_{0}\left(\mathbf{r},\mathbf{r}^{\prime},\omega\right)=-\nabla 
g_{0}\left(\mathbf{r},\mathbf{r}^{\prime},\omega\right)$.
Finally, the term that relates the magnetization to the magnetic field (magnetic field induction)
can be treated in a similar way as the one for the electric field
Green's function, the only difference is that we also have to use
integration by parts for the Laplacian term. The steps to treat this
term are exactly the same as the ones to treat the $\nabla^{\prime}\nabla^{\prime}$
term in Sec. \ref{subsec:Free-space} and we obtain
\begin{multline}
\int_{\backslash 
V_{\delta}(\mathbf{r})}d^{3}\mathbf{r}^{\prime}g_{0}\left(\mathbf{r},\mathbf{r}^{\prime},
\omega\right)\left(-\nabla^{
\prime2}+\nabla^{\prime}\nabla^{\prime}\right)\mathbf{M}_{f}(\mathbf{r}^{\prime},\omega)=\\
=\int_{\backslash 
V_{\delta}(\mathbf{r})}d^{3}\mathbf{r}^{\prime}\left(-\nabla^{\prime2}+\nabla^{\prime}\nabla^{\prime}\right)g_
{0}
\left(\mathbf{r},\mathbf{r}^{\prime},\omega\right)\mathbf{M}_{f}(\mathbf{r}^{\prime},\omega)+L_{V_{\delta}}
\mathbf{M}_{f}(\mathbf{r},\omega)-\overleftrightarrow{L}_{V_{\delta}}\cdot\mathbf{M}_{f}(\mathbf{r},\omega),
\end{multline}
where $\overleftrightarrow{L}_{V_{\delta}}$ is given by Eq.~(\ref{eq:depolarization_dyadic})
and $L_{V_{\delta}}=\text{Tr}\left(\overleftrightarrow{L}_{V_{\delta}}\right)$,
see Eq.~(\ref{eq:solid_angle}). This quantity is just the solid
angle of excluded volume $V_{\delta}$ centered at $\mathbf{r}^{\prime}=\mathbf{r}$
divided by $4\pi$, which is equals $1$ for any surface (see Appendix
\ref{sec:GF_Helmholtz}). We also point out that for $\mathbf{r}^{\prime}\neq\mathbf{r}$
we have 
$-\nabla^{\prime2}g_{0}\left(\mathbf{r},\mathbf{r}^{\prime},\omega\right)=k_{n}^{2}g_{0}\left(\mathbf{r},
\mathbf{r}^{\prime},\omega\right)$.
These results allow us to write 
\begin{align}
\label{eq:LS-E}
\mathbf{E}(\mathbf{r},\omega) & =\mathbf{E}_{0}(\mathbf{r},\omega)+\omega^{2}\mu_{n}\mu_{0}\int 
d^{3}\mathbf{r}^{\prime}\overleftrightarrow{G_0}^{EE}\left(\mathbf{r},\mathbf{r}^{\prime},
\omega\right)\cdot\mathbf{P}_{f}(\mathbf{r}^{\prime},\omega)\nonumber \\
 & +\omega\mu_{n}\mu_{0}k_{n}\int 
d^{3}\mathbf{r}^{\prime}\overleftrightarrow{G_0}^{EM}\left(\mathbf{r},\mathbf{r}^{\prime},
\omega\right)\cdot\mathbf{M}_{f}(\mathbf{r}^{\prime},\omega),\\
\label{eq:LS-B}
\mathbf{B}(\mathbf{r},\omega) & =\mathbf{B}_{0}(\mathbf{r},\omega)-\omega\mu_{n}\mu_{0}k_{n}\int 
d^{3}\mathbf{r}^{\prime}\overleftrightarrow{G_0}^{ME}\left(\mathbf{r},\mathbf{r}^{\prime},
\omega\right)\cdot\mathbf{P}_{f}(\mathbf{r}^{\prime},\omega)\nonumber \\
 & +\mu_{n}\mu_{0}k_{n}^{2}\int 
d^{3}\mathbf{r}^{\prime}\overleftrightarrow{G_0}^{MM}\left(\mathbf{r},\mathbf{r}^{\prime},
\omega\right)\cdot\mathbf{M}_{f}(\mathbf{r}^{\prime},\omega),
\end{align}
where we have the electric field and magnetic field Green's functions
\begin{align}
\overleftrightarrow{G_0}^{EE}\left(\mathbf{r},\mathbf{r}^{\prime},\omega\right) & 
=\text{P}.\text{V}._{V_{\delta}}\left[\overleftrightarrow{I}+\frac{1}{k_{n}^{2}}\nabla\nabla\right]g_{0}
\left(\mathbf{r},\mathbf{r}^{\prime},\omega\right)-\frac{1}{k_{n}^{2}}\overleftrightarrow{L}_{V_{\delta}}
\delta\left(\mathbf{r}-\mathbf{r}^{\prime}\right),\\
\overleftrightarrow{G_0}^{MM}\left(\mathbf{r},\mathbf{r}^{\prime},\omega\right) & 
=\text{P}.\text{V}._{V_{\delta}}\left[\overleftrightarrow{I}+\frac{1}{k_{n}^{2}}\nabla\nabla\right]g_{0}
\left(\mathbf{r},\mathbf{r}^{\prime},\omega\right)+\frac{1}{k_{n}^{2}}\left(\overleftrightarrow{I}
-\overleftrightarrow{L}_{V_{\delta}}\right)\delta\left(\mathbf{r}-\mathbf{r}^{\prime}\right),\label{eq:GF_MM}
\end{align}
and we have the mixed Green's functions defined as
\begin{equation}
\overleftrightarrow{G_0}^{EM}\left(\mathbf{r},\mathbf{r}^{\prime},\omega\right)=\overleftrightarrow{G_0}^{
ME}\left(\mathbf{r},\mathbf{r}^{\prime},\omega\right)=\text{P}.\text{V}._{V_{\delta}}\left[\begin{array}{ccc}
0 & -\partial_{z} & \partial_{y}\\
\partial_{z} & 0 & -\partial_{x}\\
-\partial_{y} & \partial_{x} & 0
\end{array}\right]\frac{i}{k_{n}}g_{0}\left(\mathbf{r},\mathbf{r}^{\prime},\omega\right).
\end{equation}
These describe magnetoelectric effects, which can be important when the nanoparticle 
sits on a substrate~\cite{renormalized-polarizability-2}.
The dyadic 
$\left(\overleftrightarrow{I}-\overleftrightarrow{L}_{V_{\delta}}\right)\delta\left(\mathbf{r}-\mathbf{r}^{
\prime}\right)$
in Eq.~(\ref{eq:GF_MM}) can be interpreted as a demagnetization
term. For the case for a spherically symmetric excluded volume, we
have 
$\left(\overleftrightarrow{I}-\overleftrightarrow{L}_{V_{\delta}}\right)\delta\left(\mathbf{r}-\mathbf{r}^{
\prime}\right)=\overleftrightarrow{I}2/3\delta\left(\mathbf{r}-\mathbf{r}^{\prime}\right)$.
The factor of $2/3$ is well known as being the demagnetization factor
of a spherical particle \cite{Jackson}, however, to the best of our
knowledge, this term  has not been discussed in the literature before
in the context of application of Green's functions to electromagnetic
problems. Correctly taking this term into account is essentially to
describe self-field effects in the magnetization of a particle 
(analogous to the self-field effects in the depolarization in the (electric-only) case considered before).  

For the case of nanoparticle characterized by a permetivity $\epsilon_{\text{np}}$
and permeability $\mu_{\text{np}}$, the free polarization and magnitization
densities inside the nanoparticle volume read 
\begin{align}
\mathbf{P}_{f}(\mathbf{r},\omega) & 
=\mathbf{P}_{\text{np}}(\mathbf{r},\omega)-\mathbf{P}_{n}(\omega)=\epsilon_{0}\left(\epsilon_{\text{np}}
-\epsilon_{n}\right)\mathbf{E}(\mathbf{r},\omega),\\
\mathbf{M}_{f}(\mathbf{r},\omega) & 
=\mathbf{M}_{\text{np}}(\mathbf{r},\omega)-\mathbf{M}_{n}(\omega)=\mu_{0}^{-1}\left(\mu_{n}^{-1}-\mu_{\text{np
}}^{-1}\right)\mathbf{B}(\mathbf{r},\omega),
\end{align}
where $\mathbf{P}_{\text{np}}(\mathbf{r},\omega)$ and $\mathbf{P}_{n}(\omega)$
are the polarization densities of the nanoparticle and host medium,
and $\mathbf{M}_{\text{np}}(\mathbf{r},\omega)$ and $\mathbf{M}_{n}(\omega)$
are their densities. We used the linear consititutive relations 
$\mathbf{P}_{n}(\mathbf{r},\omega)=\epsilon_{0}\left(\epsilon_{n}-1\right)\mathbf{E}(\mathbf{r},\omega)$
and $\mathbf{M}_{n}(\mathbf{r},\omega)=\mu_{0}^{-1}\left(1-\mu_{n}^{-1}\right)\mathbf{B}(\mathbf{r},\omega)$
\footnote{The previous relation between the magnetic field induction and the magnetization follows from: since $\mathbf{H}=\mathbf{B}/\mu_{0}-\mathbf{M}$ and
	$\mathbf{M}=\chi\mathbf{H}=\chi(\mathbf{B}/\mu_{0}-\mathbf{M})$, where $\chi$ is the magnetic susceptibility, then
	 $\mathbf{M}(1+\chi)=\chi\mathbf{B}/\mu_{0}\Leftrightarrow\mathbf{M}=\frac{\chi}{1+\chi}\mathbf{B}/\mu_{0}\Leftrightarrow\mathbf{M}=\frac{\chi+1-1}{1+\chi}\mathbf{B}/\mu_{0}\Leftrightarrow\mathbf{M}=(1-\mu^{-1})\mathbf{B}/\mu_{0}$. The same reasoning provides the relation between the polarization and the electric field.}. 
Inserting the two previous equations in Eqs.~(\ref{eq:LS-E})  and (\ref{eq:LS-B}), 
we obtain 
\begin{align}
\mathbf{E}(\mathbf{r},\omega) & 
=\mathbf{E}_{0}(\mathbf{r},\omega)+\omega^{2}\mu_{n}\mu_{0}\epsilon_{0}\left(\epsilon_{\text{np}}-\epsilon_{n}
\right)\int_{V}d^{3}\mathbf{r}^{\prime}\overleftrightarrow{G}_{0}^{EE}\left(\mathbf{r},\mathbf{r}^{\prime},
\omega\right)\cdot\mathbf{E}(\mathbf{r}^{\prime},\omega)\nonumber \\
 & 
+\omega\mu_{n}k_{n}\left(\mu_{n}^{-1}-\mu_{\text{np}}^{-1}\right)\int_{V}d^{3}\mathbf{r}^{\prime}
\overleftrightarrow{G}_{0}^{EM}\left(\mathbf{r},\mathbf{r}^{\prime},\omega\right)\cdot\mathbf{B}(\mathbf{r}^{
\prime},\omega),\label{eq:L-S-E}\\
\mathbf{B}(\mathbf{r},\omega) & 
=\mathbf{B}_{0}(\mathbf{r},\omega)+\mu_{n}k_{n}^{2}\left(\mu_{n}^{-1}-\mu_{\text{np}}^{-1}\right)\int_{V}d^{3}
\mathbf{r}^{\prime}\overleftrightarrow{G}_{0}^{MM}\left(\mathbf{r},\mathbf{r}^{\prime},
\omega\right)\cdot\mathbf{B}(\mathbf{r}^{\prime},\omega),\nonumber \\
 & 
-\omega\mu_{n}\mu_{0}k_{n}\epsilon_{0}\left(\epsilon_{\text{np}}-\epsilon_{n}\right)\int_{V}d^{3}\mathbf{r}^{
\prime}\overleftrightarrow{G}_{0}^{ME}\left(\mathbf{r},\mathbf{r}^{\prime},\omega\right)\cdot\mathbf{E}
(\mathbf{r}^{\prime},\omega),\label{eq:L-S-B}
\end{align}
The set of coupled equations (\ref{eq:L-S-E}) and (\ref{eq:L-S-B}) are the Lippmann-Schwinger equations for 
electromagnetic scattering. Solving them, we can obtain the  electric and magnetic fields scattered
by the nanoparticle.

\subsection{Weyl's  or angular spectrum representation of magnetic and mixed Green's functions}

Now we will see what is the Weyl's (or angular spectrum) representation
of the magnetic and mixed Green's functions. The magnetic Green's
function is almost the same as the electric Green's function, the
only difference being the different the additional 
$\overleftrightarrow{I}\delta\left(\mathbf{r}-\mathbf{r}^{\prime}\right)/k_{n}^{2}$
self-field term, which is isotropic and independent of the chosen
excluded volume. Therefore, we can write 
\begin{equation}
\overleftrightarrow{G_0}^{MM}\left(\mathbf{p}_{\parallel},z,z^{\prime},\omega\right)=\hat{e}_{s}\hat{e}_{s}
\frac{i}{2\beta_{n}}e^{i\beta_{n}\left|z-z^{\prime}\right|}+\hat{e}_{p,n}^{\pm}\hat{e}_{p,n}^{\pm}\frac{i}{
2\beta_{n}}e^{i\beta_{n}\left|z-z^{\prime}\right|}+\frac{1}{k_{n}^{2}}\left(\overleftrightarrow{I}-\hat{e}_{z}
\hat{e}_{z}\right)\delta\left(z-z^{\prime}\right).\label{eq:GMM_in_Weyl_representation}
\end{equation}
We point out that the demagnetization term $\left(\overleftrightarrow{I}-\hat{e}_{z}\hat{e}_{z}\right)$
was previously obtained in Ref.~\cite{Sipe-1987}. As for the mixed
Green's function, their Weyl's representation can be obtained by making
the replacements: $g_{0}\left(\mathbf{r},\mathbf{r}^{\prime},\omega\right)\rightarrow 
g_{0}\left(\mathbf{p}_{\parallel},z,z^{\prime},\omega\right)$,
$\left(\partial_{x},\partial_{y}\right)\rightarrow i\mathbf{p}_{\parallel}$
and $\partial_{z}\rightarrow\pm i\beta_{n}$ for $z\gtrless z^{\prime}$.
Therefore, we obtain 
\begin{equation}
\overleftrightarrow{G_0}^{EM}\left(\mathbf{p}_{\parallel},z,z^{\prime},\omega\right)=\overleftrightarrow{G}_
{0}^{ME}\left(\mathbf{p}_{\parallel},z,z^{\prime},\omega\right)=\frac{1}{k_{n}}\left[\begin{array}{ccc}
0 & \sigma\beta_{n} & -p_{y}\\
-\sigma\beta_{n} & 0 & p_{x}\\
p_{y} & -p_{x} & 0
\end{array}\right]\frac{i}{2\beta_{n}}e^{i\beta_{n}\left|z-z^{\prime}\right|},
\end{equation}
where $\sigma=\pm1$ for $z\gtrless z^{\prime}$. As for the electric
and the magnetic Green's functions, the mixed Green's functions in
the Weyl representation can also be written in terms of the $s$--and $p$--polarization vectors. 
It is straightforward to verify that 
\begin{equation}
\frac{1}{k_{n}}\left[\begin{array}{ccc}
0 & \sigma\beta_{n} & -p_{y}\\
-\sigma\beta_{n} & 0 & p_{x}\\
p_{y} & -p_{x} & 0
\end{array}\right]=\hat{e}_{p,n}^{\pm}\hat{e}_{s}-\hat{e}_{s}\hat{e}_{p,n}^{\pm},
\end{equation}
which allows us to write 
\begin{equation}
\overleftrightarrow{G_0}^{EM}\left(\mathbf{p}_{\parallel},z,z^{\prime},\omega\right)=\overleftrightarrow{G}_
{0}^{ME}\left(\mathbf{p}_{\parallel},z,z^{\prime},\omega\right)=\left[\hat{e}_{p,n}^{\pm}\hat{e}_{s}-\hat{e}_{
s}\hat{e}_{p,n}^{\pm}\right]\frac{i}{2\beta_{n}}e^{i\beta_{n}\left|z-z^{\prime}\right|}.
\end{equation}
This representation is useful, as it allows for a simple interpretation
of the emitted fields generated by the electric and magnetic dipoles
in terms of $s$-- and $p$--polarized electromagnetic waves.

If we are interested in the problem of scattering at a planar interface
between two dielectric media with $\epsilon_{1}$ for $z>0$, and $\epsilon_{2}$
for $z<0$, we can construct reflected and transmitted Green's functions
expressed in terms of reflection and transmission coefficients, as
done previously for $\overleftrightarrow{G_0}^{EE}\left(\mathbf{p}_{\parallel},z,z^{\prime},\omega\right)$.
However, some care must be taken in what the polarization vectors
mean in Green's function, considering that the polarization of an electromagnetic
field is usually defined by the polarization of the $\mathbf{E}$
field. The quantity $\overleftrightarrow{G_0}^{EM}\left(\mathbf{p}_{\parallel},z,z^{\prime},\omega\right)$
gives us the electric field generated by a point magnetic dipole located
at $z^{\prime}$. Therefore, the reflected and transmitted Green's
functions are constructed in the same way as for 
$\overleftrightarrow{G_0}^{EE}\left(\mathbf{p}_{\parallel},z,z_{0},\omega\right)$
 and for $z_{0}>0$ we obtain
\begin{align}
\overleftrightarrow{G_r}^{EM}\left(\mathbf{p}_{\parallel},z,z_{0},\omega\right) & 
=r_{p}\frac{i}{2\beta_{1}}\hat{e}_{p,1}^{+}\hat{e}_{s}e^{i\beta_{1}(z+z_{0})}-r_{s}\frac{i}{2\beta_{1}}\hat{e}
_{s}\hat{e}_{p,1}^{-}e^{i\beta_{1}(z+z_{0})},\\
\overleftrightarrow{G_t}^{EM}\left(\mathbf{p}_{\parallel},z,z_{0},\omega\right) & 
=t_{p}\frac{i}{2\beta_{1}}\hat{e}_{p,2}^{-}\hat{e}_{s}e^{-i\beta_{2}z}e^{i\beta_{1}z_{0}}-t_{s}\frac{i}{
2\beta_{1}}\hat{e}_{s}\hat{e}_{p,1}^{-}e^{-i\beta_{2}z}e^{i\beta_{1}z_{0}}.
\end{align}

For the magnetic Green's functions, 
$\overleftrightarrow{G}^{MM}\left(\mathbf{p}_{\parallel},z,z^{\prime},\omega\right)$
and $\overleftrightarrow{G}^{ME}\left(\mathbf{p}_{\parallel},z,z^{\prime},\omega\right)$,
we must take into account that these describe a field $\mathbf{B}$
generated by, respectively, a point magnetic and electric dipole.
For electric and magnetic dipoles, $\mathbf{d}_{0}$ and $\mathbf{m}_{0}$,
located at $z_{0}$, the primary magnetic field emitted for $z_{0}>z>0$
is given by
\begin{align}
\mathbf{B}_{0}\left(\mathbf{p}_{\parallel},z,\omega\right) & 
=\mu_{1}\mu_{0}k_{n}^{2}\overleftrightarrow{G}_{0}^{MM}\left(\mathbf{p}_{\parallel}z,z_{0},
\omega\right)\cdot\mathbf{m}_{0}-\omega\mu_{1}\mu_{0}k_{n}\overleftrightarrow{G}_{0}^{ME}\left(\mathbf{p}_{
\parallel}z,z_{0},\omega\right)\cdot\mathbf{d}_{0}\nonumber \\
 & 
=B_{0,s}e^{i\beta_{1}\left|z-z_{0}\right|}\hat{e}_{s}+B_{0,p}e^{i\beta_{1}\left|z-z_{0}\right|}\hat{e}_{p,1}^{
-},
\end{align}
with
\begin{align}
\label{eq:B0s}
B_{0,s} & 
=\mu_{n}\mu_{0}k_{1}^{2}\frac{i}{2\beta_{1}}\left(\hat{e}_{s}\cdot\mathbf{m}_{0}\right)+\omega\mu_{1}\mu_{0}k_
{1}\frac{i}{2\beta_{n}}\left(\hat{e}_{p,1}^{-}\cdot\mathbf{d}_{0}\right),\\
\label{eq:B0p}
B_{0,p} & 
=\mu_{n}\mu_{0}k_{1}^{2}\frac{i}{2\beta_{1}}\left(\hat{e}_{p,1}^{-}\cdot\mathbf{m}_{0}\right)-\omega\mu_{1}
\mu_{0}k_{1}\frac{i}{2\beta_{n}}\left(\hat{e}_{s}\cdot\mathbf{d}_{0}\right).
\end{align}
The corresponding electric field can be obtained from Maxwell's equations
as 
$\mathbf{E}_{0}(\mathbf{p}_{\parallel},z,\omega)=-\omega^{-1}\mathbf{p}^{\pm}_n\times\mathbf{B}_{0}(\mathbf{p}_{\parallel},z,\omega)$. More explicitly (for $z>0$)
we have for the primary field
\begin{equation}
\mathbf{E}_{0}\left(\mathbf{p}_{\parallel},z,\omega\right)=v_{1}B_{0,s}e^{i\beta_{1}\left|z-z_{0}\right|}\hat{
e}_{p,1}^{-}-v_{1}B_{0,p}e^{i\beta_{1}\left|z-z_{0}\right|}\hat{e}_{s}.
\end{equation}
This primary electric field is scattered by the interface at $z=0$,
giving origin to a reflected field for $z>0$, which reads
\begin{equation}
\mathbf{E}_{r}\left(\mathbf{p}_{\parallel},z>0,\omega\right)=r_{p}v_{1}B_{0,s}e^{i\beta_{1}\left(z+z_{0}
\right)}\hat{e}_{p,1}^{+}-r_{s}v_{1}B_{0,p}e^{i\beta_{1}\left(z+z_{0}\right)}\hat{e}_{s},
\label{eq:Er}
\end{equation}
and to a transmitted field for $z<0$
\begin{equation}
\mathbf{E}_{t}\left(\mathbf{p}_{\parallel},z<0,\omega\right)=t_{p}v_{1}B_{0,s}e^{-i\beta_{2}z}e^{i\beta_{1}z_{
0}}\hat{e}_{p,2}^{-}-t_{s}v_{1}B_{0,p}e^{-i\beta_{2}z}e^{i\beta_{1}z_{0}}\hat{e}_{s}.
\label{eq:Et}
\end{equation}
The corresponding magnetic fields are [using 
Faraday's law applied to Eqs. (\ref{eq:Er}) and (\ref{eq:Et}). For example: 
	if $\mathbf E=E_0\hat e_{p,n}$ then
	$ i\omega\mathbf{B}=\nabla\times \mathbf {E}=i\mathbf{p}^-_n\times \hat{e}^-_{p,n} E_0=ik_n\hat{e}_sE_0$, where $k_n=\vert p^-_n\vert$ and 
	$E_0$ is the amplitude of the $s-$component of the field.]
	 given by 
\begin{align}
\label{eq:Br}
\mathbf{B}_{r}\left(\mathbf{p}_{\parallel},z>0,\omega\right) & 
=r_{p}B_{0,s}e^{i\beta_{1}\left(z+z_{0}\right)}\hat{e}_{s}+r_{s}B_{0,p}e^{i\beta_{1}\left(z+z_{0}\right)}\hat{
e}_{p,1}^{+},\\
\label{eq:Bt}
\mathbf{B}_{t}\left(\mathbf{p}_{\parallel},z<0,\omega\right) & 
=t_{p}\frac{v_{1}}{v_{2}}B_{0,s}e^{-i\beta_{2}z}e^{i\beta_{1}z_{0}}\hat{e}_{s}+t_{s}\frac{v_{1}}{v_{2}}B_{0,p}
e^{-i\beta_{2}z}e^{i\beta_{1}z_{0}}\hat{e}_{p,2}^{-}.
\end{align}
From the above equations, (\ref{eq:Br}) and (\ref{eq:Bt}), we can obtain, after replacing Eqs. (\ref{eq:B0s}) and (\ref{eq:B0p}) in  Eqs. (\ref{eq:Br}) and (\ref{eq:Bt}), the reflected and transmitted
magnetic Green's functions, which are given by
\begin{align}
\overleftrightarrow{G}_{r}^{MM}\left(\mathbf{p}_{\parallel},z,z^{\prime},\omega\right) & 
=\frac{i}{2\beta_{1}}e^{i\beta_{1}\left(z+z_{0}\right)}\left[r_{p}\hat{e}_{s}\hat{e}_{s}+r_{s}\hat{e}_{p,1}^{+
}\hat{e}_{p,1}^{-}\right],\\
\overleftrightarrow{G}_{r}^{ME}\left(\mathbf{p}_{\parallel},z,z^{\prime},\omega\right) & 
=\frac{i}{2\beta_{1}}e^{i\beta_{1}\left(z+z_{0}\right)}\left[r_{p}\hat{e}_{s}\hat{e}_{p,1}^{-}-r_{s}\hat{e}_{p
,1}^{+}\hat{e}_{s}\right],\\
\overleftrightarrow{G}_{t}^{MM}\left(\mathbf{p}_{\parallel},z,z^{\prime},\omega\right) & 
=\frac{i}{2\beta_{1}}e^{-i\beta_{2}z}e^{i\beta_{1}z_{0}}\frac{v_{1}}{v_{2}}\left[t_{p}\hat{e}_{s}\hat{e}_{s}
+t_{s}\hat{e}_{p,2}^{-}\hat{e}_{p,1}^{-}\right],\\
\overleftrightarrow{G}_{t}^{ME}\left(\mathbf{p}_{\parallel},z,z^{\prime},\omega\right) & 
=\frac{i}{2\beta_{1}}e^{-i\beta_{2}z}e^{i\beta_{1}z_{0}}\frac{v_{1}}{v_{2}}\left[t_{p}\hat{e}_{s}\hat{e}_{p,1}
^{-}-t_{s}\hat{e}_{p,2}^{-}\hat{e}_{s}\right].
\end{align}
Notice that the Fresnel reflection and transmission coefficients are defined
for the electric field. With the last four equations we conclude the development of the formalism for the calculation of the renormalized polarizability \cite{renormalized-polarizability-2}
of a nanoparticle possessing both  electric and magnetic dipoles.

\section{Conclusions}

In this paper we have studied the influence of two plasmonic structures
in the effective polarizability 
of a nanoparticle made of either a metal (with a dispersionless bare polarizability) or a polar dielectric or semiconductor (with a resonant polarizability due to polar optical phonons).
 The two studied structures are a continuous graphene sheet
and a plasmonic graphene-based grating. In both cases a significant
enhancement of the imaginary part of the polarizability has been observed.
The two media  possess plasmonic resonances which, however, occur
at different frequencies. In the particular case of the grating, the
resonance is tunable in two different ways: by adjusting the gate voltage
and by changing the geometric parameters of the grating. In this case,
it is possible to scan the resonance from the THz to  the mid-IR, whereas
for the continuous graphene sheet  the resonance is always in the
THz for the currently achieved values of electronic doping using a
gate.
The approach  
pursued
 here was to model the nanoparticle by a point
like dipole. 
The main motivation for this approach lies in its ability to make analytic progress.
However, in real systems, one has a finite-size particle
which can be modeled as an assemble of many point like dipoles. 
These
are determined by the coupled dipole equations \cite{Draine_1994}. In this case, the
particle, even a spherical-one, has other multipole resonances that
can couple to the incoming radiation and contribute to the extinction
cross-section (see Appendix \ref{sec:Derivation-rate}). The two lowest
multipoles,  besides the electric dipole, are the magnetic dipole
and the electric quadrupole. It can be shown numerically that for
semiconductor nanoparticles such as spheres, cubes, pyramids, disks, and cylinders, the
the extinction cross-section has a strong magnetic-dipole resonance
\cite{magnetic-dipole,renormalized-polarizability-2,nl2012_Bozh,scirep12}
(we note, however, that for semiconductor nanoparticles, if we consider interband transitions, that is,  exciton resonances that are characteristic of semiconductos, the relevance of higher multipole resonances depends much more on the underlying band structure than on the shape). 
The formalism used in this paper to describe
the renormalization of the electric dipole resonances can be extended
to include the problem of a magnetic dipole resonance \cite{renormalized-polarizability-2}, as we have seen in the previous section.
The contribution to the extinction cross section of the magnetic dipole
is given by $\sigma_{{\rm ext}}^{{\rm m}}=\frac{\omega\mu}{2S_{{\rm inc}}}\Im([\mathbf{H}_{0}^{\ast}(\mathbf{r_{0}})\cdot\mathbf{m}(\mathbf{r_{0}})]$,
where $S_{{\rm inc}}$ is the power per unit area of the incoming
radiation and $\mathbf{m}(\mathbf{r_{0}})=\bar{\alpha}_{{\rm MM}}\text{\ensuremath{\mathbf{H}}}_{0}(\mathbf{r_{0}})$,
with $\overleftrightarrow{\alpha}_{{\rm MM}}$ the effective magnetic polarizability
of the particle and $\text{\ensuremath{\mathbf{H}}}_{0}$ the incoming
magnetic field. The effective magnetic polarizability can be derived
as done before for the electric dipole case. To that end, we 
will need
the dyadic magnetic Green's function which can be obtained from writing
the wave equation for the magnetic field using the procedure outlined
in Sec. \ref{sec:Inclusion-of-magnetic}. This study will
be pursued in a forthcoming paper.

\acknowledgments{
	B. Amorim received funding from the European Union’s Horizon 2020 research and innovation programme under grant agreement No 706538.
	N. M. R. Peres and M. I. Vasilevsiy acknowledge useful discussions with Jaime Santos and
	support from the European Commission through the project ``Graphene-Driven
	Revolutions in ICT and Beyond\textquotedbl{} (Ref. No. 696656) and
	the Portuguese Foundation for Science and Technology (FCT) in the
	framework of the Strategic Financing UID/FIS/04650/2013. 
	P.~A.~D.~Gon\c{c}alves acknowledges fruitful discussions with N.~Asger~Mortensen. 
	The Center for Nanostructured Graphene is funded by the 
	Danish National Research Foundation (project DNRF103).}


\authorcontributions{BA and NMRP did calculations and contributed to the writing of the paper. All authors contributed equaliy to the discussion of the results and writing of the paper.}

\conflictsofinterest{The authors declare no conflict of interest} 

\abbreviations{The following abbreviations are used in this manuscript:\\

\noindent 
\begin{tabular}{@{}ll}
SNOM & Scanning Nearfield Optical Microscope\\
RHS & Right-hand-side\\
NP   & Nanoparticle
\end{tabular}}

\appendixsections{multiple} 

\appendix

\section{Derivation of the wave equation}\label{sec:Derivation-of-wave_equation}

Let us start revising the basics of electromagnetic theory writing
Maxwell's equations for a homogeneous medium of relative dielectric
permittivity $\epsilon_{n}$ and relative permeability $\mu_{n}$:

\begin{align}
\nabla\times\mathbf{E}(\mathbf{r},t) & =-\frac{\partial\mathbf{B}(\mathbf{r},t)}{\partial 
t},\label{eq:maxwell1}\\
\nabla\times\mathbf{H}(\mathbf{r},t) & =\frac{\partial\mathbf{D}(\mathbf{r},t)}{\partial 
t}+\mathbf{j}_{f}(\mathbf{r},t),\label{eq:mawell2}\\
\nabla\cdot\mathbf{D}(\mathbf{r},t) & =\rho_{f}(\mathbf{r},t),\label{eq:maxwell3}\\
\nabla\cdot\mathbf{B}(\mathbf{r},t) & =0,\label{eq:maxwell4}
\end{align}
The free current density $\mathbf{j}_{f}(\mathbf{r},t)$ (current
per unit volume) and the free charge density $\rho_{f}(\mathbf{r},t)$
(charge per unit volume) are linked via the continuity equation: 
\begin{equation}
\nabla\cdot\mathbf{j}_{f}(\mathbf{r},t)+\frac{\partial\rho_{f}(\mathbf{r},t)}{\partial 
t}=0.\label{eq:continuity_eq}
\end{equation}
By free, we mean those currents that are not already taken into account by
the polarization an magnetization densities included in electric displacement,
$\mathbf{D}(\mathbf{r},t)$, and in the magnetic strength field, $\mathbf{H}(\mathbf{r},t)$.
The connection between the displacement and electric fields, and between
the magnetic induction and magnetic strength fields is given by (for
linear media) 
\begin{align}
\mathbf{D}(\mathbf{r},t) & =\epsilon_{n}\epsilon_{0}\mathbf{E}(\mathbf{r},t),\label{eq:displacement_vector}\\
\mathbf{H}(\mathbf{r},t) & =\mu_{n}^{-1}\mu_{0}^{-1}\mathbf{B}(\mathbf{r},t),\label{eq:magnetic_field}
\end{align}
where $\epsilon_{n}$ and $\mu_{n}$ are the medium relative permittivity
and permeability. Taking the curl of equation (\ref{eq:maxwell1})
and using equation (\ref{eq:mawell2}) we obtain the wave equation
for the electric field
\begin{align}
\nabla\times\nabla\times\mathbf{E}(\mathbf{r},t)+\frac{1}{v_{n}^{2}}\frac{\partial^{2}\mathbf{E}(\mathbf{r},t)
}{\partial t^{2}} & =-\mu_{n}\mu_{0}\frac{\partial\mathbf{j}_{f}(\mathbf{r},t)}{\partial 
t}\label{eq:wave_equation}
\end{align}
where $v_{n}=\sqrt{1/(\mu_{n}\mu_{0}\epsilon_{n}\epsilon_{0})}$ is
the speed of light in the medium. Let us now consider harmonic fields
with a time dependence $e^{-i\omega t}$. In this case the wave equation
reads

\begin{equation}
\nabla\times\nabla\times\mathbf{E}(\mathbf{r},\omega)-\frac{\omega^{2}}{v_{n}^{2}}\mathbf{E}(\mathbf{r},
\omega)=i\omega\mu_{n}\mu_{0}\mathbf{j}_{f}(\mathbf{r},\omega).\label{eq:wave_equation_b}
\end{equation}
Taking the curl of equation (\ref{eq:mawell2}) we find a wave equation
for the magnetic induction 
\begin{equation}
\nabla\times\nabla\times\mathbf{B}(\mathbf{r},t)+\frac{1}{v_{n}^{2}}\frac{\partial^{2}\mathbf{B}(\mathbf{r},t)
}{\partial t^{2}}=\mu_{n}\mu_{0}\nabla\times\mathbf{j}_{f}(\mathbf{r},t).
\end{equation}
Considering a harmonic time dependence of the fields and of the current
it follows 
\begin{equation}
\nabla\times\nabla\times\mathbf{B}(\mathbf{r},\omega)-\frac{\omega^{2}}{v_{n}^{2}}\mathbf{B}(\mathbf{r},
\omega)=\mu_{n}\mu_{0}\nabla\times\mathbf{j}_{f}(\mathbf{r},\omega).\label{eq:wave_equation_B_field}
\end{equation}
It is possible to rewrite Eqs.~(\ref{eq:wave_equation_b}) and (\ref{eq:wave_equation_B_field})
as inhomogeneous Helmholtz equations. In order to do that, we make
use of the identity 
$\nabla\times\nabla\times\mathbf{v}=\nabla\left(\nabla\cdot\mathbf{v}\right)-\nabla^{2}\mathbf{v}$
and write Eqs.~(\ref{eq:wave_equation_b}) and (\ref{eq:wave_equation_B_field})
as
\begin{align}
-\nabla^{2}\mathbf{E}(\mathbf{r},\omega)-\frac{\omega^{2}}{v_{n}^{2}}\mathbf{E}(\mathbf{r},\omega) & 
=i\omega\mu_{n}\mu_{0}\mathbf{j}_{f}(\mathbf{r},\omega)-\nabla\left(\nabla\cdot\mathbf{E}(\mathbf{r},
\omega)\right),\\
-\nabla^{2}\mathbf{B}(\mathbf{r},\omega)-\frac{\omega^{2}}{v_{n}^{2}}\mathbf{B}(\mathbf{r},\omega) & 
)=\mu_{n}\mu_{0}\nabla\times\mathbf{j}_{f}(\mathbf{r},\omega)-\nabla\left(\nabla\cdot\mathbf{B}(\mathbf{r},
\omega)\right).
\end{align}
Next, we use Eq.~(\ref{eq:maxwell4}) to write $\nabla\cdot\mathbf{B}(\mathbf{r},\omega)=0$,
and Eqs.~(\ref{eq:mawell2}) and (\ref{eq:displacement_vector})
to write 
$\nabla\cdot\mathbf{E}(\mathbf{r},\omega)=\epsilon_{n}^{-1}\epsilon_{0}^{-1}\rho_{f}(\mathbf{r},\omega)$.
Using the continuity equation (\ref{eq:continuity_eq}), the free
charge density can be written in terms of the free current density,
as $\rho_{f}(\mathbf{r},\omega)=\nabla\cdot\mathbf{j}_{f}(\mathbf{r},\omega)/(i\omega)$.
Therefore, we have that the electric and magnetic fields obbey the
inhomogeneous Helmholtz equations
\begin{align}
-\nabla^{2}\mathbf{E}(\mathbf{r},\omega)-\frac{\omega^{2}}{v_{n}^{2}}\mathbf{E}(\mathbf{r},\omega) & 
=i\omega\mu_{n}\mu_{0}\left[\mathbf{j}_{f}(\mathbf{r},\omega)+\frac{v_{n}^{2}}{\omega^{2}}
\nabla\left(\nabla\cdot\mathbf{j}_{f}(\mathbf{r},\omega)\right)\right],\label{eq:Helmholtz_E_field}\\
-\nabla^{2}\mathbf{B}(\mathbf{r},\omega)-\frac{\omega^{2}}{v_{n}^{2}}\mathbf{B}(\mathbf{r},\omega) & 
=\mu_{n}\mu_{0}\nabla\times\mathbf{j}_{f}(\mathbf{r},\omega).\label{eq:Helmholtz_B_field}
\end{align}
The solution to these equations can be expressed in terms of the Green's
function for the Helmholtz equation. 

\section{Green's function for the Helmholtz equation}\label{sec:GF_Helmholtz}

The inhomogeneous scalar Helmholtz equation for a field $\phi(\mathbf{r})$
and non-homogeneous source term $h(\mathbf{r})$ is given by
\begin{equation}
\left[-\nabla^{2}-k_{n}^{2}\right]\phi(\mathbf{r})=j(\mathbf{r}).\label{eq:inhomogeneous_Helmholtz}
\end{equation}
The solution for this equation can be expressed in terms of the Helmholtz
Green's function as
\begin{equation}
\phi(\mathbf{r})=\phi_{0}(\mathbf{r})+\int_{\backslash 
V_{\delta}(\mathbf{r})}d^{3}\mathbf{r}^{\prime}g_{0}\left(\mathbf{r},\mathbf{r}^{\prime
}
\right)j(\mathbf{r}^{\prime}),\label{eq:Helmholtz_integral_sol}
\end{equation}
where $\phi_{0}(\mathbf{r})$ is a particular solution of the Helmholtz
equation, $\left[-\nabla^{2}-k_{n}^{2}\right]\phi_{0}(\mathbf{r})=0$,
$g_{0}\left(\mathbf{r},\mathbf{r}^{\prime}\right)$ is the retarded
Helmholtz Green's function, which is given by Eq.~(\ref{eq:Helmholtz_greenfunction})
(we have dropped the frequency argument) and $\int_{\backslash V_{\delta}\left(\mathbf{r}\right)}$
excludes an infinitesimal volume enclosing the point $\mathbf{r}^{\prime}=\mathbf{r}$.
The goal of this appendix is to prove that Eq.~(\ref{eq:Helmholtz_integral_sol})
with $g_{0}\left(\mathbf{r},\mathbf{r}^{\prime}\right)$ given by
Eq.~(\ref{eq:Helmholtz_greenfunction}) is indeed a solution of the
inhomogeneous Helmholtz equation. In order to do that we will first
solve Eq.~(\ref{eq:inhomogeneous_Helmholtz}) by decomposing it in
terms of Fourier components, allowing a simple derivation of 
$g_{0}\left(\mathbf{r},\mathbf{r}^{\prime}\right)$.
However, that derivation does not clarify how the integration in Eq.~(\ref{eq:Helmholtz_integral_sol})
should be performed. Therefore, we will also prove that Eq.~(\ref{eq:Helmholtz_integral_sol})
solves the inhomogeneous Helmholtz equation by direct substitution. 

Writing all the fields in Fourier components
\begin{equation}
\phi(\mathbf{r})=\int\frac{d^{3}\mathbf{p}}{\left(2\pi\right)^{3}}e^{i\mathbf{p}\cdot\mathbf{r}}\phi(\mathbf{p
}),
\end{equation}
and similarly for $j(\mathbf{r})$, Eq.~(\ref{eq:inhomogeneous_Helmholtz})
becomes an algebraic equation with solution given by 
$\phi(\mathbf{p})=g_{0}\left(\mathbf{p}\right)j(\mathbf{p})$,
where
\begin{equation}
g_{0}\left(\mathbf{p}\right)=\frac{1}{p^{2}-k_{n}^{2}}.\label{eq:GF_Helmholtz_fourier}
\end{equation}
is the Helmholtz Green's function in Fourier space. Inverting the
Fourier transform, we can write
\begin{equation}
\phi(\mathbf{r})=\int 
d^{3}\mathbf{r}^{\prime}g_{0}\left(\mathbf{r},\mathbf{r}^{\prime}\right)j(\mathbf{r}^{\prime}),
\end{equation}
with \cite{scalar-green-function,Duffy}
\begin{equation}
g_{0}\left(\mathbf{r},\mathbf{r}^{\prime}\right)=\int\frac{d^{3}\mathbf{p}}{\left(2\pi\right)^{3}}\frac{e^{
i\mathbf{p}\cdot\left(\mathbf{r}-\mathbf{r}^{\prime}\right)}}{p^{2}-k_{n}^{2}}.
\end{equation}
In order to evaluate this integral we make the replacement $k_{n}\rightarrow k_{n}+i0^{+}$in
order to obtain a retarded response function. The angular integration
is easily performed and yields
\begin{equation}
g_{0}\left(\mathbf{r},\mathbf{r}^{\prime}\right)=\frac{1}{2\pi\left|\mathbf{r}-\mathbf{r}^{\prime}\right|}
\int_{-\infty}^{+\infty}\frac{dp}{2\pi 
i}\frac{pe^{ip\left|\mathbf{r}-\mathbf{r}^{\prime}\right|}}{p^{2}-\left(k_{n}+i0^{+}\right)^{2}}.
\end{equation}
The remaining integration over $p$ can be performed using contour
integration techniques, by closing the contour on the upper complex
half-plane and collecting the residue at $p=k_{n}+i0^{+}$ and obtain
Eq.~(\ref{eq:Helmholtz_greenfunction}). Notice that in order to
close the integral into the upper half-plane we must assume that $\mathbf{r}-\mathbf{r}^{\prime}\neq0$.
Next we will prove by direct substitution that Eq.~(\ref{eq:Helmholtz_integral_sol})
with the Green's function given by the above equation solves the inhomogeneous
Helmholtz equation Eq.~(\ref{eq:inhomogeneous_Helmholtz}). By doing
so, we will check that the integration in Eq.~(\ref{eq:Helmholtz_integral_sol})
actually excludes the point $\mathbf{r}=\mathbf{r}^{\prime}$. 

The crucial point in proving that Eq.~(\ref{eq:Helmholtz_integral_sol})
actually solves the inhomogeneous Helmholtz equation is to notice
that the integration region over $\mathbf{r}^{\prime}$ is actually
a function of $\mathbf{r}$. Therefore, we can write
\begin{equation}
\nabla\int_{\backslash 
V_{\delta}\left(\mathbf{r}\right)}d^{3}\mathbf{r}^{\prime}g_{0}\left(\mathbf{r},\mathbf{r}^{\prime}
\right)j(\mathbf{r}^{\prime})=-\int_{\partial 
V_{\delta}\left(\mathbf{r}\right)}d^{2}\mathbf{r}^{\prime}\mathbf{n}^{\prime}g_{0}\left(\mathbf{r},\mathbf{r}^
{\prime}\right)j(\mathbf{r}^{\prime})+\int_{\backslash 
V_{\delta}\left(\mathbf{r}\right)}d^{3}\mathbf{r}^{\prime}\nabla 
g_{0}\left(\mathbf{r},\mathbf{r}^{\prime}\right)j(\mathbf{r}^{\prime}),
\end{equation}
where $\partial V_{\delta}\left(\mathbf{r}\right)$ is the surface
of the infinitesimal volume centered at $\mathbf{r}^{\prime}=\mathbf{r}$
and $\mathbf{n}^{\prime}$ is a outwards pointing unit vector, normal
to $\partial V_{\delta}\left(\mathbf{r}\right)$. In the limit of
an infinitesimal volume element the boundary term in the above equation
vanishes, as $\delta$ is the characteristic linear size of $V_{\delta}\left(\mathbf{r}\right)$,
then we have $d^{2}\mathbf{r}^{\prime}\sim\delta^{2}$ while 
$g_{0}\left(\mathbf{r},\mathbf{r}^{\prime}\right)\sim1/\delta$.
Therefore, we can write
\begin{align}
\nabla^{2}\int_{\backslash 
V_{\delta}\left(\mathbf{r}\right)}d^{3}\mathbf{r}^{\prime}g_{0}\left(\mathbf{r},\mathbf{r}^{\prime}
\right)j(\mathbf{r}^{\prime}) & =\nabla\cdot\int_{\backslash 
V_{\delta}\left(\mathbf{r}\right)}d^{3}\mathbf{r}^{\prime}\nabla 
g_{0}\left(\mathbf{r},\mathbf{r}^{\prime}\right)j(\mathbf{r}^{\prime})\nonumber \\
 & =-\int_{\partial V_{\delta}\left(\mathbf{r}\right)}d^{2}\mathbf{r}^{\prime}\mathbf{n}^{\prime}\cdot\nabla 
g_{0}\left(\mathbf{r},\mathbf{r}^{\prime}\right)j(\mathbf{r}^{\prime})+
\int_{V_{\delta}\left(\mathbf{r}\right)}d^{3}\mathbf{r}^{\prime}\nabla^{2}g_{0}\left(\mathbf{r},\mathbf{r}^{
\prime}
\right)j(\mathbf{r}^{\prime}).
\end{align}
The boundary term now actually gives a finite contribution. To see
that, first we notice that in the limit of an infinitesimal volume
we have that $\mathbf{r}^{\prime}\rightarrow\mathbf{r}$ and therefore
we can replace $j(\mathbf{r}^{\prime})\rightarrow j(\mathbf{r})$.
Next we notice that
\begin{equation}
\nabla 
g_{0}\left(\mathbf{r},\mathbf{r}^{\prime}\right)=-\frac{e^{ik_{n}\left|\mathbf{r}-\mathbf{r}^{\prime}\right|}}
{4\pi\left|\mathbf{r}-\mathbf{r}^{\prime}\right|^{2}}\left(1-ik_{n}\left|\mathbf{r}-\mathbf{r}^{\prime}
\right|\right)\frac{\mathbf{r}-\mathbf{r}^{\prime}}{\left|\mathbf{r}-\mathbf{r}^{\prime}\right|},
\end{equation}
such that we can approximate for $\mathbf{r}^{\prime}\rightarrow\mathbf{r}$
\begin{equation}
\nabla 
g_{0}\left(\mathbf{r},\mathbf{r}^{\prime}\right)\simeq\frac{1}{4\pi\left|\mathbf{r}^{\prime}-\mathbf{r}
\right|^{2}}\frac{\mathbf{r}^{\prime}-\mathbf{r}}{\left|\mathbf{r}^{\prime}-\mathbf{r}\right|}.
\label{eq:del_g0_limit}
\end{equation}
Therefore we can write
\begin{equation}
\int_{\partial V_{\delta}\left(\mathbf{r}\right)}d^{2}\mathbf{r}^{\prime}\mathbf{n}^{\prime}\cdot\nabla 
g_{0}\left(\mathbf{r},\mathbf{r}^{\prime}\right)j(\mathbf{r}^{\prime})=L_{V_{\delta}}j(\mathbf{r}),
\end{equation}
where
\begin{equation}
L_{V_{\delta}}=\int_{\partial 
V_{\delta}\left(\mathbf{r}\right)}\frac{d^{2}\mathbf{r}^{\prime}}{4\pi}\frac{\mathbf{n}^{\prime}
\cdot\left(\mathbf{r}^{\prime}-\mathbf{r}\right)}{\left|\mathbf{r}^{\prime}-\mathbf{r}\right|^{3}}.
\label{eq:solid_angle}
\end{equation}
Therefore, it we act directly with $\left[-\nabla^{2}-k_{n}^{2}\right]$
on Eq.~(\ref{eq:Helmholtz_integral_sol}) we obtain
\begin{align}
\left[-\nabla^{2}-k_{n}^{2}\right]\phi(\mathbf{r}) & 
=\left[-\nabla^{2}-k_{n}^{2}\right]\phi_{0}(\mathbf{r})+\left[-\nabla^{2}-k_{n}^{2}\right]\int_{\backslash 
V_{\delta}\left(\mathbf{r}\right)}d^{3}\mathbf{r}^{\prime}g_{0}\left(\mathbf{r},\mathbf{r}^{\prime}
\right)j(\mathbf{r}^{\prime})\nonumber \\
 & =\int_{\backslash 
V_{\delta}\left(\mathbf{r}\right)}d^{3}\mathbf{r}^{\prime}\left[-\nabla^{2}-k_{n}^{2}\right]g_{0}\left(\mathbf
{r},\mathbf{r}^{\prime}\right)j(\mathbf{r}^{\prime})+L_{V_{\delta}}j(\mathbf{r}).
\end{align}
The first term in the first line is zero, since $\phi_{0}(\mathbf{r})$
is a solution of the homogeneous Helmholtz equation, while the first
term in the last line is zero, since 
$\left[-\nabla^{2}-k_{n}^{2}\right]g_{0}\left(\mathbf{r},\mathbf{r}^{\prime}\right)=0$
for $\mathbf{r}\neq\mathbf{r}^{\prime}$. Therefore, we obtain
\begin{equation}
\left[-\nabla^{2}-k_{n}^{2}\right]\phi(\mathbf{r})=L_{V_{\delta}}j(\mathbf{r}).
\end{equation}
Next we notice that the quantity $L_{V_{\delta}}$ is actually $1$
and is independent of the shape of the excluded volume, $V_{\delta}\left(\mathbf{r}\right)$.
First we notice that $L_{V_{\delta}}$ is actually just the solid
angle of the surface $\partial V_{\delta}\left(\mathbf{r}\right)$
that encloses the point $\mathbf{r}$ divided by $4\pi$. For a sphere
the solid angle is $4\pi$ and therefore $L_{\text{Sphere}_{\delta}}=1$.
For any other surface, we notice that the solid angle is just the
flux of the vector field
\begin{equation}
\mathbf{F}(\mathbf{r}^{\prime})=\frac{\left(\mathbf{r}^{\prime}-\mathbf{r}\right)}{\left|\mathbf{r}^{\prime}
-\mathbf{r}\right|^{3}},
\end{equation}
which satisfies $\nabla^{\prime}\cdot\mathbf{F}(\mathbf{r}^{\prime})=0$
for $\mathbf{r}^{\prime}\neq\mathbf{r}$. Therefore, for any volume
$V_{\delta}(\mathbf{r})$ enclosing the point $\mathbf{r}$, we can
consider a enclosed sphere $\text{Sphere}_{\delta}(\mathbf{r})$ and
then write
\begin{equation}
L_{V_{\delta}}=\int_{\partial\text{Sphere}_{\delta}(\mathbf{r})}\frac{d^{2}\mathbf{r}^{\prime}}{4\pi}\mathbf{n
}^{\prime}\cdot\mathbf{F}(\mathbf{r}^{\prime})+\int_{\partial\left[V_{\delta}(\mathbf{r})-\text{Sphere}_{
\delta}(\mathbf{r})\right]}\frac{d^{2}\mathbf{r}^{\prime}}{4\pi}\mathbf{n}^{\prime}\cdot\mathbf{F}(\mathbf{r}^
{\prime}).
\end{equation}
Since in the volume $V_{\delta}(\mathbf{r})-\text{Sphere}_{\delta}(\mathbf{r})$
(the volume $V_{\delta}(\mathbf{r})$ excluding the enclosing sphere)
the field $\mathbf{F}(\mathbf{r}^{\prime})$ is regular, we can use
the divergence theorem and obtain that the last term of the above
equation is zero. 

Therefore, we have obtained not only the explicit form of the Helmholtz
Green's function but have also shown that Eq.~(\ref{eq:Helmholtz_integral_sol})
is a solution of the inhomogeneous Helmholtz equation, emphasizing
the role played by the excluded volume in the integration of 
$g_{0}\left(\mathbf{r},\mathbf{r}^{\prime}\right)$.
In the case of a vector Helmholtz equation, we can use a Cartesian
basis and then use the scalar Helmholtz equation for each of the components. 

\section{Scattering and extinction cross-sections, and the role of the self-field}\label{sec:Derivation-rate}

Let us provide here a short derivation of equation (\ref{eq:rate})
using classical arguments. The power radiated by a current with time
harmonic dependence is given by Ohm's law \cite{Novotny}

\begin{equation}
\frac{dW}{dt}=-\frac{1}{2}\int 
d\mathbf{r}\Re\left[\mathbf{j}_{t}^{\dagger}(\mathbf{r},\omega)\cdot\mathbf{E}(\mathbf{r},\omega)\right].
\end{equation}
For a dipole, the current reads 
$\mathbf{j}_{t}(\mathbf{r},\omega)=-i\omega\mathbf{d}_{0}\delta(\mathbf{r}-\mathbf{r}_{0})$
which implies

\begin{equation}
\frac{dW}{dt}=\frac{\omega}{2}\Im\left[\mathbf{d}_{0}^{\dagger}\cdot\mathbf{E}(\mathbf{r}_{0},\omega)\right].
\end{equation}
Recalling that 
$\mathbf{E}(\mathbf{r},\omega)=\mu\mu_{0}\omega^{2}\overleftrightarrow{G}(\mathbf{r}-\mathbf{r}_{0},
\omega)\cdot\mathbf{d}_{0}$,
we obtain

\begin{align}
\frac{dW}{dt}=\mu\mu_{0}\omega^{2}\frac{\omega}{2}\Im\left[\mathbf{d}_{0}^{\dagger}\cdot\overleftrightarrow{G}
(0,\omega)\cdot\mathbf{d}_{0}\right]=\mu\mu_{0}\omega^{2}\frac{\omega}{2}\mathbf{d}_{0}^{\dagger}\cdot\Im\left
[\overleftrightarrow{G}(0,\omega)\right]\cdot\mathbf{d}_{0},\label{eq:scatt_cross_section}
\end{align}
where we have assumed the dipole moment real. Equation (\ref{eq:scatt_cross_section})
can also be interpreted as the scattering cross section of the scatterer
represented by the dipole when divided by the incoming power per unit
area. and replacing the dipole moment by its expression in terms of
the incoming field and polarizability. Since the energy of a photon
is $E=\hbar\omega$, the transition rate is obtained from the previous
equation as

\begin{equation}
\frac{1}{\tau_{{\rm 
cl}}}=\frac{1}{\hbar\omega}\frac{dW}{dt}=\mu\mu_{0}\omega^{2}\frac{1}{2\hbar}\mathbf{d}_{0}^{\dagger}
\cdot\Im\left[\overleftrightarrow{G}(0,\omega)\right]\cdot\mathbf{d}_{0},
\end{equation}
a result that differs from equation (\ref{eq:rate}) by a factor of
4. For obtaining the quantum result one has to make the change $\mathbf{d}_{0}\rightarrow2\mathbf{d}_{0}$
in the classical formula, a procedure well known in the literature
\cite{factor_4}. The extinction power is given by \cite{SERS} 
\begin{equation}
P_{{\rm extinction}}=\frac{\omega}{2}\Re[i\mathbf{d}_{0}^{\dagger}\cdot\mathbf{E}_{{\rm 
ext}}(\mathbf{r}_{0})].
\end{equation}
where $\mathbf{E}_{{\rm ext}}(\mathbf{r}_{0})$ is the external field.
Writing $\mathbf{d}_{0}^{\dagger}=\alpha_{{\rm eff}}^{\dagger}\mathbf{E}_{{\rm ext}}^{\ast}(\mathbf{r}_{0})$
it follows that 
\begin{equation}
P_{\mathrm{extinction}}=\frac{\omega}{2}\Im(\alpha_{{\rm eff}})\vert\mathbf{E}_{{\rm 
ext}}(\mathbf{r}_{0})\vert^{2}
\end{equation}
where $\alpha$ is the polarizability of the particle. The radiated
or scattered power is obtained from equation (\ref{eq:scatt_cross_section})
as 
\begin{equation}
P_{{\rm scat}}=\mu\mu_{0}\frac{\omega^{3}}{2}\vert\alpha_{{\rm eff}}\vert^{2}\mathbf{E}_{{\rm 
ext}}^{\dagger}(\mathbf{r}_{0})\cdot\Im\left[\overleftrightarrow{G}(0,\omega)\right]\cdot\mathbf{E}_{{\rm 
ext}}(\mathbf{r}_{0}).
\end{equation}
The absorbed power is defined as $P_{{\rm abs}}=P_{{\rm extinction}}-P_{{\rm scat}}$
\cite{SERS}. The previous analysis shows that the extinction power
is proportional to the imaginary part of the polarizability, whereas
the scattered or radiated power is proportional to the absolute value
squared of the polarizability. The time averaged Poynting vector of the incoming field is given by $S_{{\rm 
inc}}=\frac{1}{2}\sqrt{\epsilon_{1}}\epsilon_{0}c\vert\mathbf{E}_{{\rm inc}}(\mathbf{r}_{0})\vert^{2}$
. When we identify $\mathbf{E}_{{\rm ext}}(\mathbf{r}_{0})=\mathbf{E}_{{\rm inc}}(\mathbf{r}_{0})$
the scattering and extinction cross sections follow from $\sigma_{{\rm scat}}=P_{{\rm scat}}/S_{{\rm inc}}$
and $\sigma_{{\rm extinction}}=P_{{\rm extinction}}/S_{{\rm inc}}$.

\subsection{\bf Optical theorem and the role of the self-field}

Here, we revisit the interaction between light and a dipolar nanoparticle using a slightly different approach.
Let us consider a nanoparticle in vacuum with quasi-static polarizability given by 
$\alpha_{\rm CM} = 4\pi\epsilon_0 R^3 \frac{\epsilon_{\rm np} - 1}{\epsilon_{\rm np} + 2}$ 
(dubbed Claussis-Mossoty polarizability), interacting with an incident plane wave with amplitude $E_0$ and 
time dependence of the form $e^{-i\omega t}$. 

The optical theorem dictates
\begin{equation}
 \sigma_{\rm ext} = \frac{k}{\epsilon_0} \Im \{  \alpha_{\mathrm{CM}}  \} .
 \label{eq:ext}
\end{equation}
Furthermore, the absorption cross-section may be defined as
\begin{equation}
 \sigma_{\rm abs} = \frac{P_{\rm abs}}{I_0} = 
 \frac{ \frac{1}{2} \omega \Im \{  \mathbf{d}_0 \cdot \mathbf{E}^*_0 \} }{ \frac{1}{2} \epsilon_0 c E^2_0 } .
\end{equation}
Introducing $\mathbf{d}_0 =  \alpha_{\rm CM}  \mathbf{E}_0$ in the expression above, yields
\begin{equation}
 \sigma_{\rm abs} = \frac{k}{\epsilon_0} \Im \left\{  \alpha_{\rm CM}  \right\} , \label{eq:abs}
\end{equation}
which would mean that there is no scattering---compare Eqs. (\ref{eq:ext}) and (\ref{eq:abs}). Naturally, 
this cannot be the case.

In order to solve this apparent contradiction, we shall introduce the radiation reaction or radiation damping. 
The radiation reaction force (due to the self-field) is given by~\cite{Novotny}
\begin{equation}
 \mathbf{F}_R = \frac{q^2 \mathbf{\dddot r}}{6\pi\epsilon_0 c^3} ,
\end{equation}
which implies that
\begin{equation}
 \mathbf{E}_{\rm self} = \frac{q \mathbf{\dddot r}}{6\pi\epsilon_0 c^3} ,
\end{equation}
since $\mathbf{F}_R = q \mathbf{E}_{\rm self}$. Moreover, using $\mathbf{d}_0 = q\mathbf{r}$, we obtain
\begin{equation}
 \mathbf{E}_{\rm self} = \frac{1}{6\pi\epsilon_0 c^3} \mathbf{\dddot d}_0 .
\end{equation}
We now introduce the previous expression in
\begin{align}
 \mathbf{d}_0 &= \alpha_{\rm CM} [\mathbf{E}_0 + \mathbf{E}_{\rm self}]  
 = \alpha_{\rm CM} [\mathbf{E}_0 + \frac{1}{6\pi\epsilon_0 c^3} \mathbf{\dddot d}_0 ] ,\nonumber\\
 &= \alpha_{\rm CM} [\mathbf{E}_0 + \frac{(-i\omega)^3}{6\pi\epsilon_0 c^3} \mathbf{d}_0 ] ,\nonumber\\
 \Rightarrow \mathbf{d}_0 &= \frac{\alpha_{\rm CM}}{1- i \frac{k^3}{6\pi\epsilon_0 c^3} \alpha_{\rm CM}} \mathbf{E}_0,
\end{align}
which is the same result as in our Eq. (\ref{eq:vacuum_polarizability}) using the Green's functions, and from which we identify
\begin{equation}
 \alpha_0 = \frac{\alpha_{\rm CM}}{1- i \frac{k^3}{6\pi\epsilon_0 c^3} \alpha_{\rm CM}}.
\end{equation}
Now, using the polarizability accounting for the radiation reaction in the optical theorem, 
\begin{equation}
 \sigma_{\rm ext} = \frac{k}{\epsilon_0} \Im \left\{  \alpha_0  \right\} ,
\end{equation}
and assuming that $\alpha^2_{\rm CM} \approx |\alpha_{\rm CM}|^2$ (small dissipation), the previous equation becomes

\begin{align}
 \sigma_{\rm ext} &\simeq \frac{k}{\epsilon_0} \Im \left\{  \alpha'_{\rm CM} + i \alpha''_{\rm CM} + i\frac{k^3}{6\pi\epsilon_0} |\alpha_{\rm CM}|^2 \right\} , \nonumber\\
 &= \frac{k}{\epsilon_0} \left( \alpha''_{\rm CM} + \frac{k^3}{6\pi\epsilon_0} |\alpha_{\rm CM}|^2 \right), \nonumber\\
 &= \underbrace{\frac{k}{\epsilon_0} \Im \left\{  \alpha_{\rm CM}  \right\} }_{\sigma_{\rm abs}} 
 + \underbrace{ \frac{k^4}{6\pi\epsilon^2_0} |\alpha_{\rm CM}|^2 }_{\sigma_{\rm scatt}} ,
\end{align}
which solves the ``optical theorem'' dilemma by accounting for the radiation reaction arising from the self-field. 
Notice that we have also obtained the expressions for the absorption and scattering cross-sections one typically finds in books 
on plasmonic nanoparticles. Therefore, the incorporation of the effect of the self-field is pivotal in order to obtain the 
correct result for the optical theorem in the point-dipole limit.


\reftitle{References}



\end{document}